\documentclass[preprintnumbers,twocolumn,prx,aps,showpacs,10pt,superscriptaddress,floatfix,longbibliography]{revtex4-1}
\usepackage{graphicx}
\usepackage{epstopdf}
\usepackage{amsmath,amssymb,amsthm,enumitem}
\usepackage{color}
\definecolor{LinkColor}{RGB}{199,21,133}
\definecolor{CiteColor}{RGB}{220,20,60}
\usepackage[colorlinks=true,linkcolor=blue,anchorcolor=CiteColor,citecolor=CiteColor,urlcolor=LinkColor]{hyperref}
\usepackage{type1cm}
\usepackage{times}
\usepackage{multirow}

\begin{document}

\def\scrR{\mathcal{R}}
\def\scrh{\mathcal{H}}
\def\scrf{\mathcal{F}}
\def\scrz{\mathcal{Z}}
\def\scrp{\mathcal p}
\def\scrW{\mathcal W}
\def\scrM{\mathcal M}
\def\sfM{{\mathsf M}}
\def\sfW{{\mathsf W}}
\def\bP{{\mathbf P}}
\def\scrn{\mathcal{N}}
\def\scrc{\mathcal{C}}
\def\scrd{\mathcal{D}}
\def\scrl{\mathcal{L}}

\title{High-precision Monte Carlo study of several models in the three-dimensional U(1) universality class}
\author{Wanwan Xu}
\author{Yanan Sun}
\author{Jian-Ping Lv}
\email{jplv2014@ahnu.edu.cn}
\affiliation{Anhui Key Laboratory of Optoelectric Materials Science and Technology, Key Laboratory of Functional Molecular Solids, Ministry of Education, Anhui Normal University, Wuhu, Anhui 241000, China}
\author{Youjin Deng}
\affiliation{Hefei National Laboratory for Physical Sciences at Microscale and Department of Modern Physics, University of Science and Technology of China, Hefei, Anhui 230026, China}
\affiliation{CAS Center for Excellence and Synergetic Innovation Center in Quantum Information and Quantum Physics, University of Science and Technology of China, Hefei, Anhui 230026, China}
\begin{abstract}
We present a worm-type Monte Carlo study of several typical models in the three-dimensional (3D) U(1) universality class,
which include the classical 3D XY model in the directed flow representation and its Villain version,
as well as the 2D quantum Bose-Hubbard (BH) model with unitary filling in the imaginary-time world-line representation.
From the topology of the configurations on a torus, we sample the superfluid stiffness $\rho_s$ and the dimensionless wrapping probability $R$.
From the finite-size scaling analyses of $\rho_s$ and of $R$, we determine the critical points as
$T_c ({\rm XY}) =2.201\, 844 \,1(5)$ and $T_c ({\rm Villain})=0.333\, 067\, 04(7)$ and $(t/U)_c ({\rm BH})=0.059 \, 729 \,1(8)$,
where $T$ is the temperature for the classical models, and $t$ and $U$ are respectively the hopping and on-site interaction strength for the BH model. The precision of our estimates improves significantly over that of the existing results. Moreover, it is observed that at criticality, the derivative of a wrapping probability with respect to $T$ suffers from negligible leading corrections and enables a precise determination of the correlation length critical exponent as $\nu=0.671 \, 83(18)$. In addition, the critical exponent $\eta$ is estimated as $\eta=0.038 \, 53(48)$ by analyzing a susceptibility-like quantity. We believe that these numerical results would provide a solid reference in the study of classical and quantum phase transitions in the 3D U(1) universality, including the recent development of the conformal bootstrap method.
\end{abstract}
\maketitle

\section{Introduction}
The U(1) criticality is a textbook example of phase transition and plays a crucial role in many-body phenomena ranging from vortex binding-unbinding transition~\cite{kosterlitz1973ordering}, exotic quantum phases such as superfluid (SF)~\cite{svistunov2015superfluid} and spin liquid~\cite{hermele2004pyrochlore,hermele2004stability}, emergent continuous symmetries responsible for deconfined criticality~\cite{senthil2004deconfined,senthil2004quantum} to quantum emulating~\cite{greiner2002quantum} and to relativistic gauge field theories in particle physics~\cite{faddeev2018gauge}. Hence, the quantitative aspects of the U(1) criticality are frequently a requisite.

In three dimensions, systems within the U(1) universality class have a continuous phase transition with non-trivial critical exponents. Exact results are unavailable either for the critical points or the critical exponents. Numerical~\cite{campostrini2001critical,alet2003directed,campostrini2006theoretical,burovski2006high,lan2012high,komura2014cuda} and approximate~\cite{gottlob1993critical,campostrini2000determination,kos2016precision} methodologies
have been extensively applied. Up to now, the most precise estimates of critical exponents were obtained mostly with Monte Carlo methods, including $\nu=0.671\,7(1)$~\cite{campostrini2006theoretical} and $0.671 \, 7(3)$~\cite{burovski2006high}
and $\eta=0.038 \, 1(2)$~\cite{campostrini2006theoretical}. These estimates have been extensively utilized in literature, albeit the estimate of $\nu$ differs from the celebrated experimental result $\nu=0.670 \, 9(1)$~\cite{lipa2003specific}
determined by a specific heat measurement around the finite-temperature SF transition of $^{4}$He performed in the microgravity environment of a space shuttle. The conformal bootstrap method has given a very precise determination for the critical exponents of three-dimensional (3D) Ising ($\mathbb{Z}_2$) model, yet produced less precise estimates $\nu=0.671 \, 9(11)$ and $\eta=0.038 \, 52(64)$ for the U(1) case~\cite{kos2016precision}.
A summary of estimated critical exponents $\nu$ and $\eta$ for the 3D U(1) universality class are given by Table~\ref{Litnueta}.

In this work we carry out a high-precision Monte Carlo study of three paradigmatic models in the 3D U(1) universality class,
including the classical XY and Villain models on the simple-cubic lattice and the quantum Bose-Hubbard (BH) model with unitary filling on the square lattice. The XY model is the $n=2$ case of the O($n$) vector model and is a prototype of lattice models with continuous symmetries.
It has a broad realm of physical realizations including granular superconductors and Josephson junction arrays~\cite{goldman2013percolation}.
The XY model has been extensively studied by Monte Carlo simulations, which are mostly on the spin representation and use the celebrated cluster update schemes~\cite{swendsen1987nonuniversal,wolff1989collective}.
The Villain model is a variant of the XY model. Both the XY and Villain models can be rewritten in the directed flow representation, and then be simulated by the worm algorithm~\cite{prokof2001worm}, which is very efficient in the measurement of correlation functions.
Together with numerical analytical-continuation methods, the high-precision Monte Carlo data of the two-point correlation function, obtained by worm-type simulations, have uncovered intriguing low-energy excitations and
transport properties near the critical temperature $T_c$~\cite{chen2014universal,witczak2014dynamics}.
It was observed~\cite{chen2014universal} that the precise determination of $T_c$ is crucial in these studies.
The BH model and its extensions can describe a wide variety of quantum phenomena~\cite{dutta2015non}, including the SF, Mott insulator, supersolid and spin-liquid phases. In quantum Monte Carlo (QMC) simulations, the BH model is frequently expressed in terms of the imaginary-time world-line (path-integral) representation.
In the field of cold-atom physics, the BH model has become a prominent object of state-of-the-art optical lattice emulators~\cite{greiner2002quantum,trotzky2010suppression,endres2011observation}.
At unitary filling--i.e., each lattice site is occupied by a particle on average,
the $d$-dimensional BH model exhibits the particle-hole symmetry, and the quantum phase transition between the SF and the Mott insulating phase belongs to the $(d+1)$-dimensional XY universality.

In our worm-type Monte Carlo simulations, periodic boundary conditions are applied.
From the topology of the directed flow and the world-line configurations,
we sample the SF stiffness $\rho_s$ and the wrapping probabilities $R$'s, of which the definitions will be given below.
The data of $\rho_s$ and $R$ are analyzed according to the finite-size scaling theory,
and yield mutually consistent determinations of the critical points.
The wrapping probability is observed to suffer from smaller finite-size corrections.
The final estimates of the critical temperature, which takes into account both the statistical and the systematic uncertainties,
are $T_c ({\rm XY}) =2.201 \, 844 \, 1(5)$ and $T_c({\rm Villain})=0.333\, 067\, 04(7)$.
In a similar way, the quantum critical point (QCP) of the unitary-filling BH model is determined as $(t/U)_c=0.059 \, 729 \, 1(8)$, where $t$ is the hopping amplitude between the nearest-neighboring sites and $U$ denotes the strength of on-site repulsion.
Our estimates improve significantly over the existing results; see Table~\ref{LitCP} for details.
For instance, in comparison with $(t/U)_c=0.059 \, 74(4)$ from the strong-coupling expansion~\cite{elstner1999dynamics}
and $(t/U)_c=0.059 \, 74(3)$ by QMC simulations~\cite{capogrosso2008monte},
our result of the QCP has a higher precision by a factor of more than 40.
To determine the correlation length critical exponent $\nu$, we calculate the derivative of the wrapping probability $R$ with respect to the temperature $T$ from the covariance of $R$ and energy. For the Villain model at criticality, a quantity of this type is found to exhibit negligible finite-size corrections and yields $\nu=0.671 \, 83(18)$. This estimate is consistent with the most precise Monte Carlo results $\nu=0.671 \, 7(1)$~\cite{campostrini2006theoretical} and $0.671 \, 7(3)$~\cite{burovski2006high} with a comparable precision, and suggests that the experimental determination $\nu=0.670 \, 9(1)$~\cite{lipa2003specific} and the graphics processing unit (GPU) simulation result $\nu=0.670 \, 98(16)$~\cite{lan2012high} are unlikely. In addition, we obtain the critical exponent $\eta=0.038 \, 53(48)$ from a susceptibility-like quantity, which is very close to a recent conformal boostrap estimate $\eta=0.038 \, 52(64)$~\cite{kos2016precision}.

\begin{table}
 \begin{center}
 \caption{Estimates of the critical exponents $\nu$ and $\eta$ in the 3D U(1) universality class. The method adopted and the year when the result was published are also listed. `Ref.', `RG', `Exp.', `HTE', `MC' and `CB' are the abbreviations of `reference', `renormalization group', `experiment', `high-temperature expansion', `Monte Carlo' and `conformal bootstrap', respectively.}
 \label{Litnueta}
 \begin{tabular}[t]{|c|c|c|l|l|}
   \hline
   Ref. &  Method & Year  & $\nu$& $\eta$  \\
  \hline
  \cite{gottlob1993critical}  & RG &1993&   0.662(7) & \\
  \cite{lipa1996heat}  & Exp. &1996&  0.670\,19(13)& \\
  \cite{campostrini2000determination}  & HTE &2000&  0.671\,66(55)& 0.038\,1(3)\\
  \cite{campostrini2001critical} & MC  & 2001  & 0.671\,6(5)&  0.038\,0(5)\\
  \cite{campostrini2001critical}  & MC+HTE  &2001&  0.671\,55(27) & 0.038\,0(4)\\
   \cite{lipa2003specific}  & Exp. &2003&  0.670\,9(1)& \\
   \cite{campostrini2006theoretical}  & MC+HTE  &2006&   0.671\,7(1)&  0.038\,1(2)\\
   \cite{burovski2006high} & MC &2006&   0.671\,7(3)& \\
   \cite{lan2012high}  & MC (GPU)  &  2012  &0.670\,98(16)&  \\
   \cite{lan2012high}  & MC (GPU) & 2012  &0.671\,38(11)&  \\
   \cite{komura2014cuda} & MC (GPU)  & 2014  &0.672(4) & \\
   \cite{kos2016precision} & CB  &  2016 &0.671\,9(11)& 0.038\,52(64) \\
   this work & MC  &  2019 &0.671\,83(18)& 0.038\,53(48)\\
   \hline
 \end{tabular}
 \end{center}
 \end{table}

In the remainder of this paper, we present details for the definition of the models, the methodology and the scaling analyses of numerical results. Section~\ref{Models} introduces the models addressed in this work. Section~\ref{Methodology} elaborates the methodology which contains a unified formulation of worm Monte Carlo algorithm for the XY and Villain models, the definitions of sampled quantities, and finite-size scaling ansatze. Section~\ref{re} presents Monte Carlo results and scaling analyzes. A short summary is finally given in Sec.~\ref{sum}.

\section{Models}~\label{Models}
The Hamiltonian of the XY model reads
\begin{equation}\label{HamXY}
H_{\rm XY}=- \sum\limits_{\langle{\bf r}
{\bf r'}\rangle} {\bf S_{\bf r}} \cdot {\bf S_{{\bf r'}}},
\end{equation}
where ${\bf S_{\bf r}}$=$(\cos \theta_{\bf r}, \sin \theta_{\bf r})$ denotes a planar spin vector with unit length at site ${\bf r}$ and the summation runs over pairs of nearest neighboring sites on a simple-cubic lattice. As listed in Table~\ref{LitCP}, the most recent estimates of $T_c$ (we have set $k_{\rm B}=1$ for convenience) include $2.201\,840\,5(48)$~\cite{deng2005surface}, $2.201\,831\,2(6)$~\cite{lan2012high}, $2.201\,852(1)$~\cite{lan2012high} and $2.201\,836(6)$~\cite{komura2014cuda}. Albeit these estimates are all based on Monte Carlo simulations, they are {\it not} completely consistent with each other.

\begin{table}
 \begin{center}
 \caption{Estimates of the critical temperatures for the 3D XY and Villain models and the critical hopping amplitude for the two-dimensional (2D) unitary-filling BH model. `SCE' is the abbreviation of `strong-coupling expansion'.}
 \label{LitCP}
 \begin{tabular}[t]{|c|c|c|c|l|}
  \hline
 Model & Ref. & Method & Year  &$T_c$ or $(t/U)_c$  \\
 \hline
  {\multirow{8}{*}{XY}} & \cite{gottlob1993critical} & MC &1993 &  2.201\,67(10)\\
 &\cite{Ballesteros} & MC   &1996 &  2.201\,843(19)  \\
 &\cite{Cucchieri} & MC   &2002 &  2.201\,833(19) \\
 &\cite{deng2005surface} & MC   &2005 &  2.201\,840\,5(48) \\
 & \cite{lan2012high}   & MC (GPU) & 2012  &2.201\,831\,2(6)\\
 & \cite{lan2012high}   & MC (GPU) & 2012  &2.201\,852(1)\\
 &\cite{komura2014cuda}  & MC (GPU) &2014   &2.201\,836(6) \\
 &this work  & MC  &2019  & 2.201\,844\,1(5)\\
  \hline
 {\multirow{3}{*}{Villain}} &\cite{alet2003cluster}&  MC & 2003 &  0.333\,05(5)\\
& \cite{chen2014universal}  &  MC  & 2014 &  0.333\,067\,0(2)\\
&  this work &   MC  & 2019 & 0.333\,067\,04(7) \\
   \hline
 {\multirow{3}{*}{BH}} & \cite{elstner1999dynamics} & SCE  &1993 &  0.059\,74(4) \\
 & \cite{capogrosso2008monte} & MC  & 2008 &  0.059\,74(3) \\
  & this work  &  MC  &2019  & 0.059\,729\,1(8)\\
 \hline
 \end{tabular}
 \end{center}
 \end{table}

\begin{figure}
\includegraphics[width=8cm,height=5cm]{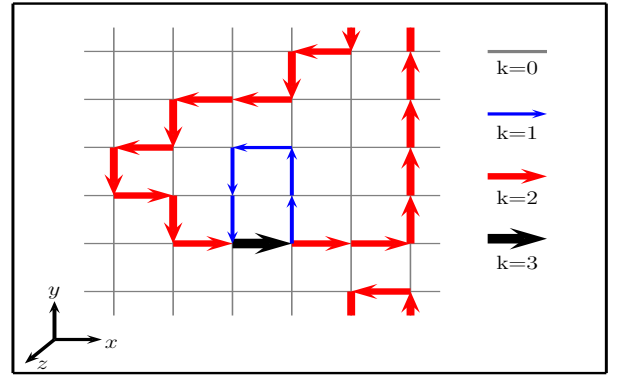}
\caption{A directed flow configuration of the XY and Villain models on a cross section of $6 \times 6 \times 6$ periodic simple-cubic lattice.}~\label{configuration}
\end{figure}

We perform a high-temperature expansion on model~(\ref{HamXY}) for the directed flow representation. We begin with the partition function
\begin{equation}\label{ZXY}
\scrz_{\rm XY}=(\frac{1}{2\pi})^N  \int e^{-\frac{H_{\rm XY}}{T}} \prod_{\bf r} d\theta_{\bf r},
\end{equation}
where $N$ is the number of lattice sites on the simple-cubic lattice and the exponential $e^{-\frac{H_{\rm XY}}{T}}$ can be expanded as
\begin{equation}\label{exp}
e^{-\frac{H_{\rm XY}}{T}}=\prod_{\langle{\bf r}{\bf r'}\rangle} e^{\frac{\cos(\theta_{\bf r}-\theta_{\bf r'})}{T}}.
\end{equation}
Next, we combine (\ref{ZXY}) and (\ref{exp}) with the Fourier transform
\begin{equation}\label{eqft}
e^{\frac{\cos(\theta_{\bf r}-\theta_{\bf r'})}{T}}=\sum^{+\infty}_{\scrc_{{\bf r}{\bf r'}}=-\infty} J_{\scrc_{{\bf r}
{\bf r'}}}(\frac{1}{T}) e^{i \scrc_{{\bf r}
{\bf r'}} (\theta_{\bf r}
-\theta_{\bf r'})},
\end{equation}
where $J_{\scrc_{{\bf r}{\bf r'}}}(\frac{1}{T})$ represents the $\scrc_{{\bf r}{\bf r'}}$th order modified Bessel function of
the first kind in variable of $\frac{1}{T}$; $J_{\scrc_{{\bf r}{\bf r'}}}(\frac{1}{T})=J_{-\scrc_{{\bf r}{\bf r'}}}(\frac{1}{T})$.
It follows that
\begin{eqnarray}\label{dr2}
&\scrz_{\rm XY}&= (\frac{1}{2\pi})^N  \sum_{\{\scrc_{{\bf r}
{\bf r'}}\}} \int  \Bigg( \prod_{\langle{\bf r}
{\bf r'}\rangle}  J_{\scrc_{{\bf r}
{\bf r'}}}(\frac{1}{T}) e^{i  \scrc_{{\bf r}
{\bf r'}} (\theta_{\bf r}
-\theta_{\bf r'}) } \Bigg) \prod_{\bf r}
 d\theta_{\bf r}
 \nonumber  \\
&= &(\frac{1}{2\pi})^N  \sum_{\{\scrc_{{\bf r}
{\bf r'}}\}}  \Bigg(\prod_{\langle{\bf r}
{\bf r'}\rangle}  J_{\scrc_{{\bf r}
{\bf r'}}}(\frac{1}{T}) \Bigg) \int \prod_{\langle{\bf r}
{\bf r'}\rangle} e^{i
 \scrc_{{\bf r}
{\bf r'}} (\theta_{\bf r}
-\theta_{\bf r'}) } \prod_{\bf r} d\theta_{\bf r}  \nonumber \\
&=& (\frac{1}{2\pi})^N  \sum_{\{\scrc_{{\bf r}
{\bf r'}}\}}   \Bigg(\prod_{\langle {\bf r}
{\bf r'}\rangle}  J_{\scrc_{{\bf r}
{\bf r'}}}(\frac{1}{T}) \Bigg) \int \prod_{\bf r}  e^{i  \scrd_{{\bf r}
} \theta_{\bf r}
 }  \prod_{\bf r} d\theta_{\bf r}. \nonumber \\ \nonumber
\end{eqnarray}
We represent the term $ \scrc_{{\bf r}
{\bf r'}} (\theta_{\bf r}
-\theta_{\bf r'})$ by a vector flow variable $\scrc_{{\bf r}
{\bf r'}}$ from ${\bf r}$ to
${\bf r'}$. As $\scrc_{{\bf r}{\bf r'}}$ is positive (negative), it means that the flow is
from ${\bf r}$ to
${\bf r'}$ (${\bf r'}$ to
${\bf r}$); $\scrc_{{\bf r}{\bf r'}}=-\scrc_{{\bf r'}{\bf r}}$. $\scrd_{{\bf r}
}=\sum_{\bf r'} \scrc_{{\bf r}{\bf r'}}$ denotes the divergence of flow at ${\bf r}$.
The inner integration would be zero unless $\Delta\scrc=0$ := ($\forall \,\, {\bf r}, \scrd_{{\bf r}}=0$).
Hence $\Delta\scrc=0$ represents the null divergence of flow over lattice sites and mimics the Kirchhoff's current law.
As a result we have
\begin{equation}\label{ZXYO}
\scrz_{\rm XY}= \sum_{\Delta\scrc=0} \prod_{\langle {\bf r}
{\bf r'}\rangle} J_{\scrc_{{\bf r}
{\bf r'}}}(\frac{1}{T}),
\end{equation}
where the summation runs over all the directed flow states with $\Delta\scrc=0$. Up to now, we have finished an exact transformation of standard XY model onto a directed flow model, for which we illustrate a configuration in Fig.~\ref{configuration}.

A variant of the standard XY model is the Villain model with partition function
\begin{equation}\label{ZVillian}
\scrz_{\rm Villain}= (\frac{1}{2\pi})^N \int \prod_{\bf r} d\theta_{\bf r} \sum\limits_{\scrl_{{\bf rr'}}=-\infty}^{+\infty} e^{-\frac{1}{2T} \sum\limits_{\langle{\bf r}
{\bf r'}\rangle}(\theta_{{\bf r'}}-\theta_{\bf r}-2 \pi \scrl_{{\bf rr'}})^2}.
\end{equation}
The $2\pi$ periodicity persists in the interaction potential of the Villain model, which is believed to capture the characteristics of XY model~\cite{cha1991universal,alet2003cluster,vsmakov2005universal,chen2014universal,witczak2014dynamics}. A high-temperature expansion~\cite{nishimori2010elements} can be performed on the Villain model to build the directed flow representation. As a result, the partition function in the directed flow space $\Delta\scrc=0$ reads
\begin{equation}
\scrz_{\rm Villain}=  \sum_{\Delta\scrc=0} \prod_{\langle {\bf r}
{\bf r'}\rangle} e^{\frac{-\scrc^2_{{\bf rr'}}}{2T}}.
\end{equation}
Note that the Villain model has a simple form of statistical weight allocation in the directed flow representation. Estimates of $T_c$ for the 3D Villain model are summarized in Table~\ref{LitCP}.

We consider the 2D unitary-filling BH model with the Hamiltonian
\begin{equation}\label{HamBH}
\hat{H}_{\rm BH}=-t \sum\limits_{\langle{\bf r}
{\bf r'}\rangle}(\hat{a}^+_{{\bf r}
} \hat{a}_{{\bf r'}}+\hat{a}^+_{{\bf r'}} \hat{a}_{{\bf r}})+\frac{U}{2} \sum\limits_{{\bf r}}\hat{n}_{{\bf r}}(\hat{n}_{{\bf r}}-1),
\end{equation}
where $\hat{a}^+_{{\bf r}}$ ($\hat{a}_{{\bf r}}$) represents the local bosonic creation (annihilation) operator at site ${\bf r}$; $\hat{n}_{{\bf r}}$=$\hat{a}^+_{{\bf r}} \hat{a}_{{\bf r}}$. The first summation runs over the pairs of nearest-neighbouring sites on a square lattice and the second one is over sites. When the particle density is fixed at integer numbers, a quantum phase transition between the compressible SF phase and the incompressible Mott insulating phase occurs by tuning the ratio $t/U$. The transition falls in the 3D U(1) universality and features a relativistic (particle-hole) symmetry with the dynamical critical exponent $z=1$. In this work we study the unitary-filling case for which the QCP has been estimated as $(t/U)_c=0.059\,74(4)$ by a high-order strong-coupling expansion~\cite{elstner1999dynamics} and $(t/U)_c=0.059\,74(3)$ by QMC simulations~\cite{capogrosso2008monte}. These two estimates have become benchmarks for the location of QCP of the 2D unitary-filling BH model (Table~\ref{LitCP}).

\section{Methodology}~\label{Methodology}
\subsection{Monte Carlo Algorithms}
Our Monte Carlo simulations of both the classical (XY and Villain) and quantum (BH) models employ worm-type algorithms. For the classical models the worm algorithm was first proposed in~\cite{prokof2001worm}. An explicit formulation of worm algorithm for the Villain model has been presented in~\cite{alet2003cluster}. It has been demonstrated~\cite{deng2007dynamic} that the worm algorithm stands out from state-of-the-art algorithms when sampling certain quantities for the 3D Ising model. For completeness, we shall formulate a worm algorithm for the XY and Villain models. As for the BH model, we use the standard worm QMC algorithm in the continuous imaginary time path-integral (world-line) representation~\cite{prokof1998exact,prokof1998worm}, for which we refrain from a detailed elaboration and refer the readers to~\cite{prokof1998exact,prokof1998worm,prokof2009worm,pollet2012recent}.

\subsubsection{Worm Algorithm for the XY model}~\label{waxy}
{\it Extending state space.---} A character of worm algorithm is enlarging state space.
It extends the original directed flow space $\Delta\scrc=0$ by including two additional degrees of freedom,
namely, two defects $I$ and $M$ individually on a site, by defining that the subspace with $I=M$ recovers exactly the
original space $\Delta\scrc=0$ and that the subspace with $I\ne M$ is exactly outside the original space.
The latter is called worm ($\scrW$) sector, where the Kirchhoff's current law does not hold for sites $I$ and $M$,
namely, $\scrd_I \neq 0, \scrd_M \neq 0$. Accordingly, the partition function for the extended state space can be
separated into two parts~\cite{lv2011worm}. The first part
\begin{equation}
\scrz_{\rm XY}=\frac{1}{N} \sum_{\Delta\scrc=0; I, M} \delta_{I,M} \prod_{\langle {\bf r}
{\bf r'}\rangle}  J_{\scrc_{{\bf r}{\bf r'}}}(\frac{1}{T})
\end{equation}
corresponds to the original partition function (\ref{ZXYO}), with $\delta$ the Kronecker delta function. The summation runs over the extended state space. The second part relates to $\scrW$ sector and reads
\begin{equation}
\scrz_\scrW=\frac{1}{N} \sum_{\Delta\scrc=0; I, M} (1-\delta_{I,M}) \prod_{\langle {\bf r}
{\bf r'}\rangle}  J_{\scrc_{{\bf r}
{\bf r'}}}(\frac{1}{T}).
\end{equation}
The Monte Carlo simulations will be performed in the extended state space with the partition function
\begin{equation}
\scrz_{\rm Ext}=\scrz_{\rm XY}+\Lambda \scrz_\scrW,
\end{equation}
where $\Lambda$ is a tunable parameter often (but not necessarily) set as $\Lambda=1$.

{\it Worm updates.---} The worm process moves $I$ or/and $M$ around lattice and updates the directed
flow configuration by changing the local flow variable through a biased random walk designed with detailed balance.

More specifically, as $I$ ($M$) moves to a neighboring site $I_\scrn$ ($M_\scrn$), the flow on edge $II_\scrn$ ($MM_\scrn$) will be
updated accordingly by adding a unit, directed flow from $I$ to $I_\scrn$ ($M_\scrn$ to $M$).
The movement repeats until $I$ and $M$ meet ($I=M$), when the original state space is hit.
Hence the movement of $I$ or $M$ serves as a step of random walk in the $\scrW$ sector or
in between the $\scrW$ sector and the original state space. The core steps are given as follows. \\
{\vspace{0.cm}\hspace{0.51cm} \textbf{Core steps of the worm algorithm for XY model}
\begin{enumerate}[leftmargin=*,labelindent=-1pt,label=\bfseries Step \arabic*.]
\item If $I=M$, randomly choose a new site $I'$ and set $I=M=I'$, ${\rm sign}(I)=1$, ${\rm sign}(M)=-1$.
\item Interchange $I \leftrightarrow M$ and ${\rm sign}(I) \leftrightarrow {\rm sign}(M)$ with probability $1/2$.
\item Randomly choose one neighboring site $I_\scrn$ of $I$.
\item Propose to move $I \rightarrow I_\scrn$ by updating the flow along edge-$II_\scrn$, $\scrc_{II_\scrn}$, to $\scrc'_{II_\scrn}$:
\begin{equation}
\scrc'_{II_\scrn}=\scrc_{II_\scrn}+{\rm sign}(I \rightarrow I_\scrn){\rm sign}(I), \nonumber
\end{equation}
where ${\rm sign}(I \rightarrow I_\scrn)=\pm 1$, parametrizing the flow direction along edge-$II_\scrn$.
\item Accept the proposal with probability
      \begin{equation}
       \bP_{\rm acc} = \min (1, \frac{J_{\scrc'_{II_\scrn}}(\frac{1}{T})}{J_{\scrc_{II_\scrn}}(\frac{1}{T})}) \nonumber
      \end{equation}
      according to the Metropolis scheme.
\end{enumerate} \vspace{0.cm}}

Monte Carlo simulations are constituted by repeating \textbf{Steps 1} to \textbf{5}. The exploration of XY model is achieved by sampling quantities as the original state space is hit. Besides, the worm process itself is informative for detecting two-point correlations. A susceptibility-like quantity $T_w$ (integral of two-point correlation) can be evaluated by the number of worm steps between subsequent hits on the original state space, which is known as returning time $\mathcal{\tau_{\rm w}}$. Accordingly, the definition of $T_w$ is given by
\begin{equation}\label{defineTw}
T_w=\langle \mathcal{\tau}_w \rangle,
\end{equation}
which scales as $T_w \sim L^{2-\eta}$ at the critical point and is useful for determining the critical exponent $\eta$.

\subsubsection{Worm Algorithm for the Villain Model}~\label{wavl}
The worm algorithm formulated in Sec.~\ref{waxy} applies to the Villain model once a substitute of \textbf{Step 5} is taken as follows.\\
{\vspace{0.cm}
\begin{enumerate}[leftmargin=*,labelindent=-1pt,label=\bfseries Step \arabic*.]
\item[\textbf{Step 5.}] Accept the proposal with probability
      \begin{equation}
       \bP_{\rm acc} = \min (1, e^{\frac{-(\scrc'^2_{II_\scrn}-\scrc^2_{II_\scrn})}{2T}}) \nonumber
      \end{equation}
      according to the Metropolis scheme.
\end{enumerate} \vspace{0.cm}}

The definition of $T_w$~(\ref{defineTw}) applies to the Villain model as well.

\subsection{Sampled Quantities}
Wrapping probability has been studied for random cluster models (including its limiting situations percolation, Ising and Potts models, etc.)~\cite{langlands1992universality,pinson1994critical,ziff1999shape,newman2001fast,arguin2002homology,martins2003percolation, wang2013bond,xu2014simultaneous,hou2019geometric}. Thus far, however, the method has not been employed for the U(1) lattice models which, as we have shown, admit a graphic representation such as directed flow representation. Moreover, the applicability of wrapping probability approach to
quantum models has not yet been addressed.

\begin{figure}
\includegraphics[width=8cm,height=3.5cm]{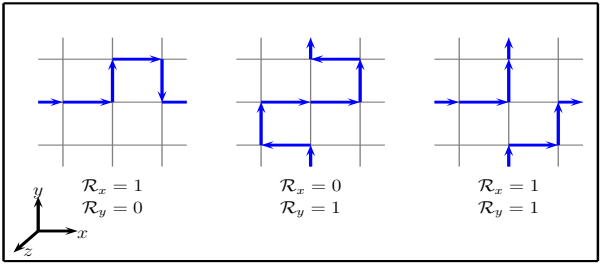}
\caption{Illustration of wrappings for directed flow configurations of the 3D XY and Villain models on a cross section of $3 \times 3 \times 3$ periodic simple-cubic lattice.}~\label{guide_wrapping}
\end{figure}

The original definition of wrapping is based on the cluster representation of percolation~\cite{langlands1992universality}. Here we extend its original definition to describe a broader class of graphic representation constituted by the directed flows (XY and Villain models) or the particle lines in the imaginary-time path-integral configuration (BH model), by exploiting the wrapping of directed flows or particle lines around lattice torus. As the flow is non-divergent, the wrapping for XY and Villain models around a certain direction can be justified by the presence of {\it net} flow passing through the perpendicular cross section. Moreover, in the present worm Monte Carlo simulations for the XY, Villain and BH models, the wrapping of directed flows or particle lines can be justified by tracking the movements of defects ($I$ and $M$). Directed flow configurations with wrapping, namely, $\scrR_{\kappa}=1$ for $\kappa$=$x$, $y$ or/and $z$, are illustrated by Fig.~\ref{guide_wrapping} for XY and Villain models.

\subsubsection{Sampled Quantities for the 3D XY and Villain Models}
Some wrapping-related quantities for 3D XY and Villain models are defined as follows.
The wrapping probabilities in the directed flow representation are given by
 \begin{eqnarray}
     R_x &=&  \langle \scrR_x \rangle = \langle \scrR_y \rangle = \langle \scrR_z \rangle, \\
     R_a &=&  1 - \langle (1-\scrR_x)(1-\scrR_y)(1-\scrR_z)\rangle, \\
     R_2 &=&  \langle \scrR_x \scrR_y (1-\scrR_z)\rangle  +\langle \scrR_y \scrR_z (1-\scrR_x)\rangle \nonumber \\  &&+\langle \scrR_z \scrR_x (1-\scrR_y)\rangle,
 \end{eqnarray}
where $R_x$, $R_a$ and $R_2$ define the probabilities that the wrapping of directed flows exists in $x$ direction, in at least one direction and in two (and only two) directions, respectively. For a wrapping observable (say $\scrR_{\kappa}$), we define its covariance with energy $\mathcal{E}$ as
 \begin{equation}
  G_{R_\kappa E}= \frac{1}{T^2} (\langle \scrR_\kappa \mathcal{E} \rangle - \langle \scrR_\kappa \rangle \langle \mathcal{E} \rangle),
 \label{eq:Rp}
 \end{equation}
which is equivalent to the derivative of $R_\kappa$ with respect to $T$.

Recall the definition of winding number on a torus. If a direction (say $\kappa$) is specified, the event {\it wrapping} (namely, $\scrR_\kappa=1$) relates to a non-zero winding number $\mathcal{W}_\kappa \ne 0$ of directed flow. The latter has a close connection to the definition of SF stiffness~\cite{pollock1987path}. We sample the SF stiffness by winding number fluctuations, which can be written as
\begin{equation}
\rho_s=\langle \mathcal{W}_x^2+\mathcal{W}_y^2+\mathcal{W}_z^2 \rangle /3L.
\end{equation}
Moreover, we estimate the derivative of $\rho_s$ with respect to $T$ by
\begin{equation}
G_{\rho_sE} = \frac{1}{3LT^2} (\langle (\mathcal{W}_x^2+\mathcal{W}_y^2+\mathcal{W}_z^2) \mathcal{E} \rangle -\langle \mathcal{W}_x^2+\mathcal{W}_y^2+\mathcal{W}_z^2 \rangle  \langle \mathcal{E}  \rangle).
\label{eq:rhosp}
\end{equation}

\subsubsection{Sampled Quantities for the 2D BH Model}
For the 2D BH model, the wrapping probabilities of particle lines in the world-line representation read
 \begin{eqnarray}
     R_x &=& \langle \scrR_x \rangle = \langle \scrR_y \rangle, \\
     R_a &=&  1 - \langle (1-\scrR_x)(1-\scrR_y)\rangle, \\
     R_2 &=& \langle \scrR_x \scrR_y\rangle,
 \end{eqnarray}
where $R_x$, $R_a$ and $R_2$ define the probabilities that the wrapping of particle lines exists in $x$ direction, in at least one direction and in two directions, respectively.

For a given spatial direction (say $\kappa$), the event {\it wrapping} (namely, $\scrR_\kappa=1$) relates to a non-zero winding number  $\mathcal{W}_\kappa \ne 0$ of particle lines. We sample the SF stiffness by winding number fluctuations through~\cite{pollock1987path}
\begin{equation}
\rho_s=\langle \mathcal{W}_x^2+\mathcal{W}_y^2 \rangle /4 t \beta.
\end{equation}

\subsection{Finite-size Scaling Ansatze}~\label{fss}
In order to formulate the finite-size scaling for the thermodynamic phase transition of the classical (XY and Villain) models and the quantum phase transition of the quantum (BH) model in a more or less unified manner, a quantum to classical mapping is required.
For the QCP of the unitary-filling BH model, the dynamic critical exponent $z=1$~\cite{fisher1989boson} has been verified extensively~\cite{cha1991universal,alet2003cluster}. In our QMC simulations of the BH model, we use the temperature contour $\beta \equiv 1/T=2L$. This treatment eliminates the variable $\beta/L^z$ in the finite-size scaling ansatz of the BH model.

Wrapping probabilities are dimensionless quantities whose finite-size scaling is described by
\begin{equation}\label{Rscaling}
R_{\kappa}=\widetilde{R}_{\kappa} (\epsilon L^{1/\nu})
\end{equation}
where $\widetilde{R}_{\kappa}$ is a scaling function, $\epsilon$ denotes the distance to critical point. We set for the XY and Villain models $\epsilon=T_c-T$ and for the BH model $\epsilon=(t/U)_c-t/U$.

Performing a Taylor's expansion and incorporating corrections to scaling, we have
 \begin{equation}\label{fit_R}
R_{\kappa}=\mathcal{Q}_0+\sum_n a_n \epsilon L^{n/\nu}+\sum_m b_m L^{-\omega_m},
\end{equation}
where $\mathcal{Q}_0$ is a somewhat universal constant, $a_n$ ($n=1,2,...$) and $b_m$ ($m=1,2,...$) are non-universal constants, and $\omega_m$ ($m=1,2,...$) refers to correction-to-scaling exponents. To estimate the critical exponent $\nu$ we analyze $G_{R_\kappa E}=\frac{d R_{\kappa}}{dT}$ which should scale as
 \begin{equation}
G_{R_\kappa E}=L^{1/\nu} {\widetilde{G}_{R_\kappa E}} (\epsilon L^{1/\nu}),
\end{equation}
and we have
 \begin{equation}\label{fit_Rp}
G_{R_\kappa E}=L^{1/\nu} (\mathcal{Q}_0+\sum_n a_n \epsilon L^{n/\nu}+\sum_m  b_m L^{-\omega_m}).
\end{equation}

The finite-size scaling of $\rho_s$ can be figured out by $\rho_s \sim \xi^{2-d-z}$~\cite{ma1986strongly,fisher1989boson} with $\xi$ the correlation length. For the present case (three space-time dimensions), we have $d+z=3$. It follows that
\begin{equation}~\label{rhosscaling}
\rho_s L=\widetilde{\rho}_s(\epsilon L^{1/\nu})
\end{equation}
and
 \begin{equation}\label{fit_Rhos}
\rho_s L=\mathcal{Q}_0+\sum_n a_n \epsilon L^{n/\nu}+\sum_m  b_m L^{-\omega_m}.
\end{equation}

The scaling form of $G_{\rho_sE}=\frac{d\rho_s}{dT}$ reads
\begin{equation}
G_{\rho_sE}=L^{(-1+\frac{1}{\nu})} \widetilde{G}_{\rho_s E} (\epsilon L^{1/\nu}),
\end{equation}
and we have
\begin{equation}\label{fit_Rhosp}
G_{\rho_sE}=L^{(-1+\frac{1}{\nu})}(\mathcal{Q}_0+\sum_n a_n \epsilon L^{n/\nu}+\sum_m b_m L^{-\omega_m}).
\end{equation}

Besides, for the XY and Villain models, one may use the following scaling form of $T_w$ to determine the critical exponent $\eta$,
\begin{equation}\label{scaling_Tw}
T_w=L^{2-\eta} \widetilde{T}_w(\epsilon L^{1/\nu}),
\end{equation}
which gives
\begin{equation}\label{fit_Tw}
T_w=L^{2-\eta}(\mathcal{Q}_0+\sum_n a_n \epsilon L^{n/\nu}+\sum_m b_m L^{-\omega_m}).
\end{equation}
In principle, an analytic background should be added up to (\ref{scaling_Tw}) and (\ref{fit_Tw}). For the present case,
this analytic background is effectively higher-order corrections compared with the correction terms taken into
account explicitly throughout this work.

\section{Numerical Results and Finite-size Scaling Analyses}~\label{re}
In this section, we present Monte Carlo results which are analyzed by performing finite-size scaling. In subsection~\ref{dtc}, firstly, the scaling behaviors of wrapping probabilities and some other quantities are explored for 3D Villain and XY models. We find that certain wrapping probabilities exhibit smaller corrections in finite-size scaling than those of the ``conventional'' quantities such as SF stiffness and susceptibility-like quantities. Universal values of critical dimensionless quantities are confirmed numerically for the XY and Villain models. Subsequently, we extend the wrapping probability approach to determine the QCP of the 2D unitary-filling BH model. For each of the Villain, XY and BH models, an unprecedentedly precise estimate of critical point is obtained. In subsection~\ref{fatc}, we determine the critical exponents $\nu$ and $\eta$ following extensive simulations at the high-precision $T_c$ of the Villain model. We find that the quantity $G_{R_xE}$ exhibits a negligible (if non-zero) amplitude of leading correction, which helps to determine a precise estimate of critical exponent $\nu$. Moreover, the critical exponent $\eta$ is estimated by analyzing the susceptibility-like quantity $T_w$.

\subsection{Determining Critical Points}\label{dtc}
\subsubsection{3D Villain Model}
We simulate the 3D Villain model on periodic $L\times L \times L$ simple-cubic lattices with linear sizes $L=16$, $24$, $32$, $64$, $128$, $256$, $384$ and $512$ for different $T$ around $T=0.333\,067$. The simulations use the worm Monte Carlo algorithm described in Sec.~\ref{wavl}. The finite-size scaling for the finite-temperature phase transition of 3D Villain model is performed by fitting the Monte Carlo data of wrapping probabilities $R_x$, $R_a$ and $R_2$ to (\ref{fit_R}), of scaled SF stiffness $\rho_s L$ to (\ref{fit_Rhos}), and of susceptibility-like quantity $T_w$ to (\ref{fit_Tw}). These fits are carried out with the least squares method. For a preferred fit, one expects that $\chi^2$ per degree of freedom (DF) $\chi^2/{\rm DF}$ is $O(1)$. Moreover, for a precaution against systematic errors brought about by the exclusion of high-order correction terms, we prefer the fits with a stability against varying the cut-off size $L_{\rm min}$, which denotes the smallest lattice of which the data are included in.

\begin{figure}
\includegraphics[width=9cm,height=7cm]{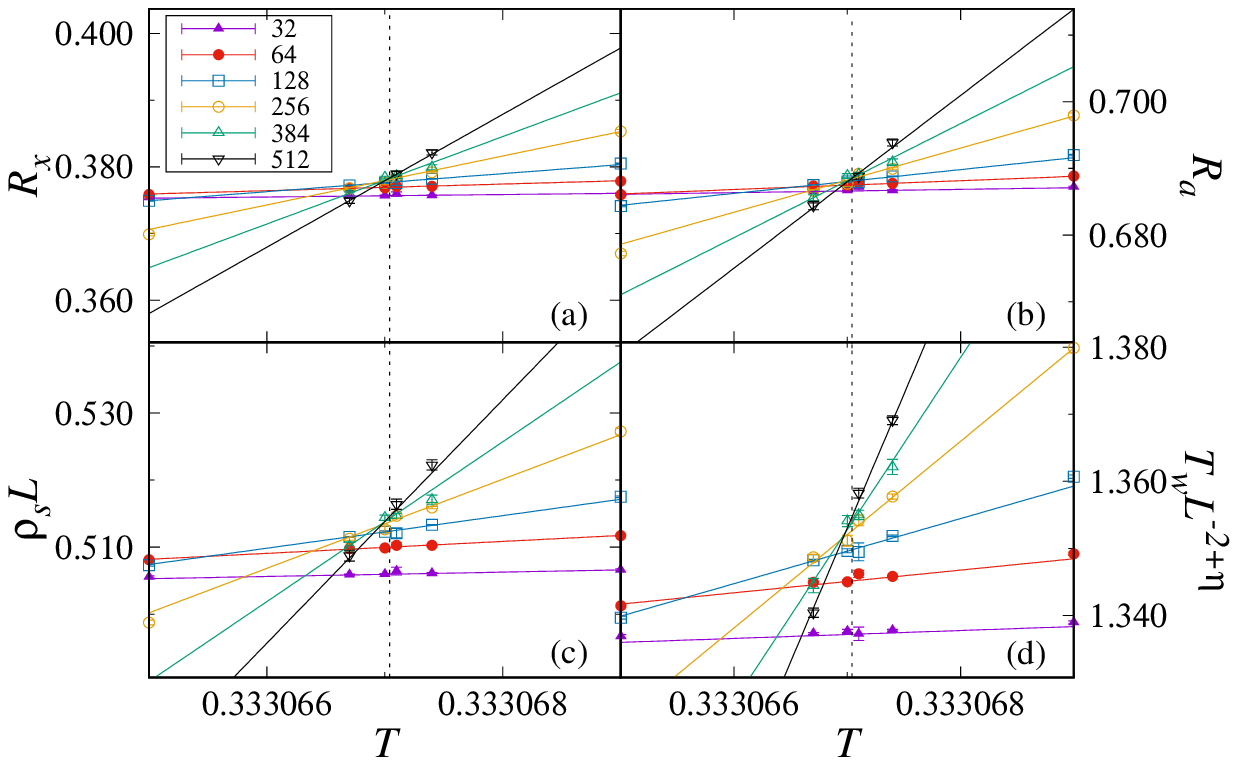}
\caption{Wrapping probabilities $R_x$ and $R_a$, scaled SF stiffness $\rho_s L$ and scaled susceptibility-like quantity $T_wL^{-2+\eta}$ versus $T$ in the neighbourhood of critical temperature for the 3D Villain model. The linear lattice sizes are $L=32$, $64$, $128$, $256$, $384$ and $512$. The symbols stand for Monte Carlo raw data, and the lines are drawn with a preferred fit with $\chi^2/{\rm DF} \approx 1$. Note that the same domain of vertical coordinate is focused on around the crossing points (estimated by $\mathcal{Q}_0$ in the fits) for the quantities to make it is fair to compare finite-size corrections. The vertical dashed lines represent our finite estimate of critical temperature $T_c=0.333\,067\,04(7)$.}~\label{Villainquantities}
\end{figure}

\begin{table*}
\begin{center}
\caption{Fits of the wrapping probabilities $R_x$, $R_a$, $R_2$ to $(\ref{fit_R})$ and the scaled SF stiffness $\rho_s L$ to $(\ref{fit_Rhos})$ for the 3D Villain model. `Qua.' is the abbreviation of `quantities'. The correction exponents $\omega_1=0.789$ and $\omega_2=1.77$ are adopted. The symbol `-' denotes the absence of the corresponding correction term in the fit.}
 \label{Tab:fit-Villian1}
 \begin{tabular}[t]{|l|l|l|l|l|l|l|l|l|}
 \hline
 Qua.   &$L_{\rm min}$ & $\chi^2/$DF  & $T_c$   & $1/\nu$   & $\mathcal{Q}_0$   & $a_1$   & $b_1$   & $b_2$   \\

 \hline
{\multirow{8}{*}{$R_x$}} &16&  27.2/26 & 0.333\,067\,035(15)&1.43(6) &0.378\,65(4) &-1.26(39)   & -0.0459(4)  & -   \\
                             &24&  24.8/22 & 0.333\,067\,038(16)&1.43(6)&0.378\,66(5) &-1.32(41)       & -0.0461(7) & -   \\
                             &32&  22.0/18 & 0.333\,067\,039(17)&1.43(6)&0.378\,67(6) &-1.31(41)   & -0.046(1)  & -   \\
                             &64&  21.0/14 & 0.333\,067\,045(22)&1.43(6)&0.378\,7(1) &-1.31(43)   & -0.047(3)  & -   \\
                             & 8&  29.1/29 & 0.333\,067\,047(16)&1.44(6)&0.378\,72(5) &-1.25(38)   & -0.0480(9)  & 0.025(5)  \\
                             &16&  26.9/25 & 0.333\,067\,041(18)&1.44(6) &0.378\,68(7) &-1.25(39)   & -0.047(2)  & 0.01(2)   \\
                             &24&  24.7/21 & 0.333\,067\,041(21)&1.43(6)&0.378\,7(1) &-1.31(41)       & -0.047(4) & 0.02(7)  \\
                             &32&  21.9/17 & 0.333\,067\,044(25)&1.43(6)&0.378\,7(2) &-1.30(41)   & -0.048(7) & 0.04(14)   \\

 \hline
 {\multirow{7}{*}{$R_a$}}
                             &24&  33.5/22 &0.333\,067\,020(18)&1.41(6)&0.688\,70(7) &-1.90(65)   & -0.031(1) & -   \\
                             &32&  23.3/18 &0.333\,067\,032(19)&1.42(6)&0.688\,79(8) &-1.78(62)   & -0.033(2)  & -   \\
                             &64&  20.7/14 &0.333\,067\,044(24)&1.42(7) &0.688\,9(2) &-1.81(65)   & -0.037(4)  & -   \\
                             & 8&  34.9/29 &0.333\,067\,058(18)&1.43(6)&0.689\,05(7) &-1.73(59)   & -0.044(1)  & 0.194(8)  \\
                             &16&  33.0/25 &0.333\,067\,046(20)&1.43(6) &0.689\,0(1) &-1.72(59)   & -0.040(3)  & 0.15(3)   \\
                             &24&  29.9/21 &0.333\,067\,050(24)&1.42(6)&0.689\,0(2) &-1.83(63)       & -0.042(6) & 0.19(10)   \\
                             &32&  22.8/17 &0.333\,067\,046(28)&1.43(6)&0.689\,0(3) &-1.76(62)   & -0.04(1) & 0.14(21)   \\

 \hline
 {\multirow{8}{*}{$R_2$}}    &16&  31.7/26 & 0.333\,067\,074(21)&1.47(9) &0.264\,10(4) &-0.81(37)   & -0.034\,7(5)  & -   \\
                             &24&  16.1/22 & 0.333\,067\,060(23)&1.47(9)&0.264\,06(5) &-0.83(39)      & -0.033\,8(8) & -   \\
                             &32&  13.5/18 & 0.333\,067\,059(25)&1.47(9)&0.264\,05(6) &-0.82(39)  & -0.034(1)  & -   \\
                             &64&  12.2/14 & 0.333\,067\,049(31)&1.46(9)&0.264\,0(1) &-0.89(43)   & -0.032(3)  & -   \\
                             & 8&  31.2/29 & 0.333\,067\,062(23)&1.47(9)&0.264\,06(5) &-0.83(38)   & -0.033\,3(9)  & -0.015(6)  \\
                             &16&  29.1/25 & 0.333\,067\,050(26)&1.47(9) &0.264\,00(8) &-0.82(38)   & -0.032(2)  & -0.04(2)   \\
                             &24&  16.0/21 & 0.333\,067\,052(31)&1.47(9)&0.264\,0(1) &-0.84(39)       & -0.032(4) & -0.03(7)   \\
                             &32&  13.0/17 & 0.333\,067\,041(36)&1.47(9)&0.263\,9(2) &-0.82(39)   & -0.029(8)  & -0.10(15)   \\
 \hline
 {\multirow{8}{*}{$\rho_s L$}} &16&  30.4/26 & 0.333\,067\,019(15)&1.45(6) &0.515\,45(6) &-2.21(67)   & -0.1452(8)  & -   \\
                             &24&  25.5/22 & 0.333\,067\,028(16)&1.44(6)&0.515\,51(8) &-2.30(70)       & -0.146(1) & -   \\
                             &32&  24.7/18 & 0.333\,067\,029(18)&1.43(6)&0.515\,5(1) &-2.38(74)  & -0.147(2)  & -   \\
                             &64&  24.0/14 & 0.333\,067\,035(22)&1.43(6)&0.515\,6(2) &-2.39(76)  & -0.149(5)  & -   \\
                             & 8&  30.0/29 & 0.333\,067\,040(16)&1.45(6)&0.515\,66(8) &-2.14(65)   & -0.152(2) & 0.078(9)  \\
                             &16&  28.5/25 & 0.333\,067\,034(18)&1.45(6) &0.515\,6(1) &-2.17(66)   & -0.150(3)  & 0.05(4)   \\
                             &24&  25.4/21 & 0.333\,067\,033(22)&1.44(6)&0.515\,6(2) &-2.28(70)      & -0.149(7) & 0.04(12)   \\
                             &32&  24.7/17 & 0.333\,067\,033(26)&1.43(6)&0.515\,6(3) &-2.38(74)   & -0.15(1)  & 0.05(24)   \\
 \hline
 \end{tabular}
 \end{center}
 \end{table*}

\begin{table*}
\begin{center}
\caption{Fits of $T_w$ to $(\ref{fit_Tw})$ for the 3D Villain model. The correction exponents $\omega_1=0.789$ and $\omega_2=1.77$ are adopted.}
 \label{Tab:fit-Villian2}
 \begin{tabular}[t]{|l|l|l|l|l|l|l|l|l|}
 \hline
  $L_{\rm min}$ & $\chi^2/$DF  & $T_c$ & $\eta$  & $1/\nu$   & $\mathcal{Q}_0$    & $a_1$   & $b_1$   & $b_2$   \\
 \hline
  16&  29.8/28 &0.333\,066\,996(17)&0.0392(2)&1.52(4) &1.360(1) &-3.06(67)   & -0.307(6)  & -   \\
  24& 24.2/24 &0.333\,067\,021(21)&0.0386(4)&1.52(4)&1.356(2) &-3.04(67)       & -0.29(1) & -   \\
  32&  21.7/20 &0.333\,067\,027(25)&0.0385(5)&1.53(4)&1.355(4) &-2.95(66)   & -0.28(2)  & -   \\
  64&  21.1/15 &0.333\,067\,027(55)&0.0385(16)&1.53(4)&1.36(1) &-2.90(66)       & -0.28(9) & -   \\
  8&  29.4/31 &0.333\,067\,040(20)&0.0380(3)&1.52(4)&1.351(2) &-3.01(66)   & -0.24(1)  & -0.40(5)  \\
  16&  26.1/27 &0.333\,067\,042(29)&0.0379(7)&1.53(4) &1.351(5) &-2.96(65)   & -0.24(4)  & -0.4(2)   \\
  24&  24.0/23 &0.333\,067\,034(39)&0.0382(11)&1.52(4)&1.353(9) &-3.03(67)       & -0.26(8)& -0.2(6)   \\
  32&  21.7/19 &0.333\,067\,027(67)&0.0385(25)&1.53(4)&1.36(2) &-2.95(66)   & -0.3(2)  & 0(2)   \\
 \hline
 \end{tabular}
 \end{center}
 \end{table*}

We quote the leading correction exponent $\omega_1=0.789(11)$ predicated by a $d=3$ loop expansion~\cite{guida1998critical} and verified by a finite-size scaling of Monte Carlo data~\cite{campostrini2006theoretical} that produced $\omega=0.785(20)$. We shall perform a test for this correction exponent in below. As for the sub-leading correction term, we adopt the correction exponent $\omega_2=1.77$~\cite{guida1998critical}. We observe that the incorporating of correction terms stabilizes fits. In Table~\ref{Tab:fit-Villian1}, we list the details of the fits for dimensionless quantities $R_x$, $R_a$, $R_2$ and $\rho_s L$. The fitting results for $T_w$ are given by Table~\ref{Tab:fit-Villian2}. It is found that the amplitudes of leading correction (namely, $|b_1|$) for these quantities differ from each other. Each of the wrapping probabilities $R_x$, $R_a$ and $R_2$ suffers from minor corrections to scaling, with $|b_1| \lessapprox 0.05$. By contrast, the ``conventional" quantity $\rho_s L$ and $T_w$ exhibit more significant corrections with $|b_1| \approx 0.15$ and $0.3$, respectively. The distinct amplitudes of finite-size corrections for the quantities can be inferred from Fig.~\ref{Villainquantities}.

\begin{figure}
\includegraphics[width=8cm,height=6.7cm]{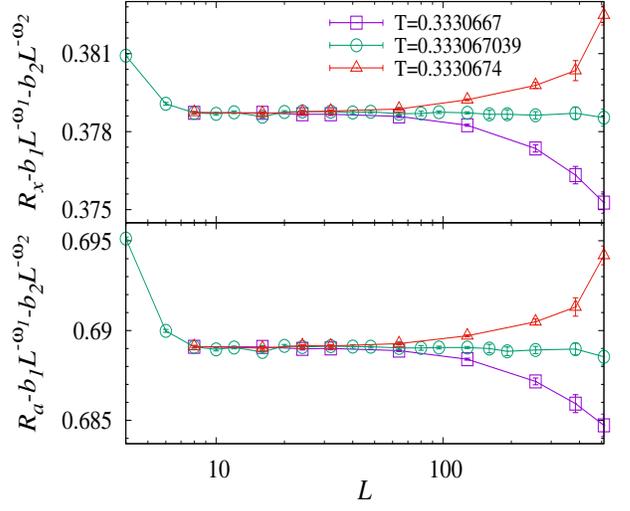}
\caption{Scaled $R_x$ and $R_a$ with correction terms versus $L$ at $T_c$ and its neighborhoods for the 3D Villain model. The corrections terms quote preferred fitting results in Table~\ref{Tab:fit-Villian1}.}~\label{Rtx}
\end{figure}

We should not trust blindly a sole fitting even though $\chi^2$/DF is close to $1$ and we do not take any individual fitting result as our final estimate. In fact, we take the medium out of all preferred fitting results of the quantities to estimate $T_c$. To be conservative, the estimated error bar measures the distance between the final estimate and the farthest bound among those indicated by individual fits. Accordingly, by using the finite-size scaling analyses of the dimensionless quantities and the quantity $T_w$ respectively presented in Tables~\ref{Tab:fit-Villian1} and~\ref{Tab:fit-Villian2}, we estimate the critical temperature as $T_c=0.333\,067\,04(7)$. As an illustration, we plot in Fig.~\ref{Rtx} the size-dependent behavior of $R_x$ and $R_a$ with correction terms at $T_c=0.333\,067\,039$ (we take $T_c=0.333\,067\,04(7)$ and keep one more decimal place) and at its neighborhoods $T=0.333\,066\,7$ and $0.333\,067\,4$. At $T_c$, the data becomes asymptotically a constant as $L \rightarrow \infty$, whereas it is either upward or downward as $T$ deviates from $T_c$. These suggest for the present size scale that $T_c=0.333\,067\,039$ is a reasonable estimate and that $T=0.333\,0667$ and $0.333\,0674$ deviate from the critical region.

\begin{figure}
\includegraphics[width=8cm,height=5.8cm]{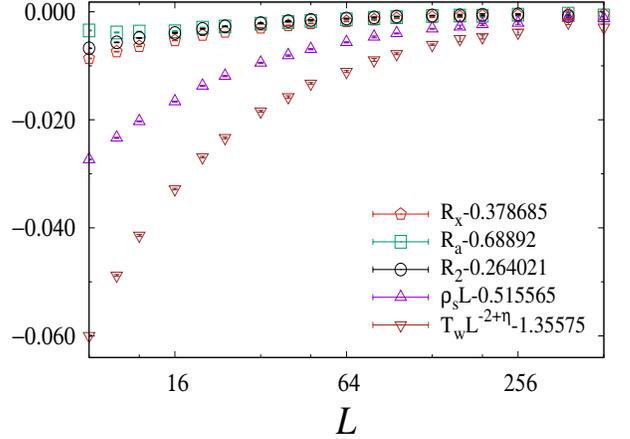}
\caption{Corrections to leading scaling revealed by $R_\kappa-\mathcal{Q}_0$ for $R_\kappa$ ($\kappa=x,a,2$), by $\rho_sL-\mathcal{Q}_0$ for $\rho_s L$, and by $T_wL^{-2+\eta}-\mathcal{Q}_0$ for $T_w$ at the estimated critical temperature $T_c=0.333\,0670\,39$ of the 3D Villain model. The parameter $\mathcal{Q}_0$ is determined from Table~\ref{Tab:fit-Villian1} for $R_\kappa$ ($\kappa=x,a,2$) and $\rho_s L$, and from Table~\ref{Tab:fit-Villian2} for $T_w$. The critical exponent $\eta$ takes our finite estimate $\eta=0.038\,53$.}~\label{corr}
\end{figure}

The relatively small finite-size corrections of $R_x$, $R_a$ and $R_2$ compared with those of $\rho_s L$ and $T_w$ can be observed in Fig.~\ref{corr}, which demonstrates the corrections to leading scaling for each of the quantities.

We determine by Table~\ref{Tab:fit-Villian1} (from the estimates of $\mathcal{Q}_0$) the critical wrapping probabilities as $R^c_x=0.378\,7(2)$, $R^c_a=0.688\,9(4)$ and $R^c_2=0.264\,0(3)$, and the critical winding number fluctuations (namely, the critical scaled SF stiffness) as $\rho^c_sL=0.515\,6(3)$.

\begin{figure}
\includegraphics[width=9cm,height=7cm]{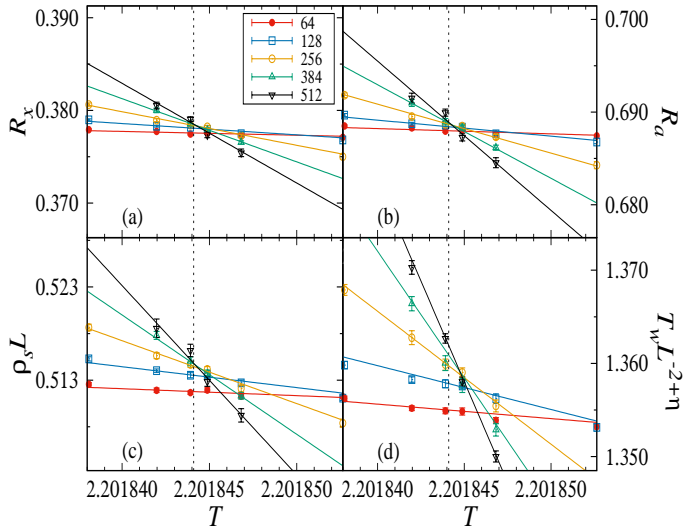}
\caption{Wrapping probabilities $R_x$ and $R_a$, scaled SF stiffness $\rho_sL$ and scaled susceptibility-like quantity $T_wL^{-2+\eta}$ versus $T$ in the neighbourhood of critical temperature for the 3D XY model. The linear lattice sizes are $L=64$, $128$, $256$, $384$ and $512$. The same domain of vertical coordinate is focused on for the quantities. The vertical dashed lines represent our finite estimate of critical temperature $T_c=2.201\,844\,1(5)$.}~\label{3dxy_tc}
\end{figure}

\subsubsection{3D XY model}
By means of the worm algorithm formulated in Sec.~\ref{waxy}, we simulate the 3D XY model (\ref{HamXY}) on periodic simple-cubic lattices with linear sizes $L=8$, $16$, $32$, $64$, $128$, $256$, $384$ and $512$ around $T=2.201\,84$. We aim at estimating the critical temperature $T_c$ of this paradigmatic model and exploring universal features of critical wrapping probabilities. We fit the quantities $R_\kappa$ ($\kappa=x,a,2$) and $\rho_s L$ to their finite-size scaling ansatze (\ref{fit_R}) and (\ref{fit_Rhos}), respectively. The fitting results are detailed in Appendix~\ref{appendix:3dxy} by which we estimate the critical temperature as $T_c=2.201\,844\,1(5)$, which is more precise than the best results available in literature, as listed in Table~\ref{LitCP}. The estimate $T_c=2.201\,844\,1(5)$ is further confirmed by the finite-size scaling analysis with a fixed $\nu$ (Appendix~\ref{appendix:3dxy}). Figure~\ref{3dxy_tc} illustrates the quantities $R_x$, $R_a$, $\rho_sL$ and $T_wL^{-2+\eta}$ as a function of $T$. Shown by Table~\ref{Tab:universalWP}, the Villain model and the XY model share universal critical values of wrapping probabilities and of winding number fluctuations.

\begin{table}
\begin{center}
\caption{Critical wrapping probabilities $R^c_x$, $R^c_a$, $R^c_2$ and critical winding number fluctuations $\rho^c_sL$ for the 3D Villain and XY models.}
 \label{Tab:universalWP}
 \begin{tabular}[t]{|c|c|c|}
 \hline
   & Villain & XY    \\
 \hline
 $R^c_x$& 0.378\,7(2)  & 0.378\,8(4) \\
 $R^c_a$& 0.688\,9(4)  & 0.688\,9(6) \\
 $R^c_2$& 0.264\,0(3)  & 0.264\,1(5) \\
 $\rho^c_sL$ & 0.515\,6(3) & 0.515\,8(9) \\
  \hline
 \end{tabular}
 \end{center}
 \end{table}

\subsubsection{2D Unitary-filling BH Model}
\begin{figure}
\includegraphics[width=9cm,height=7cm]{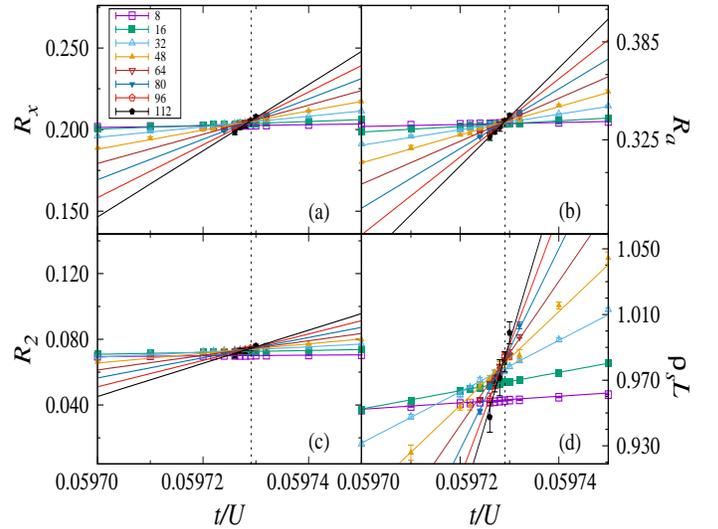}
\caption{Wrapping probabilities $R_x$, $R_a$, $R_2$ and scaled SF stiffness $\rho_s L$ in the neighborhood of QCP for the 2D unitary-filling BH model. The linear lattice sizes are $L=8$, $16$, $32$, $48$, $64$, $80$, $96$ and $112$; for each $L$, the inverse temperature is $\beta=2L$. The same domain of vertical coordinate is focused on for the quantities. The vertical dashed lines represent our finite estimate of QCP, namely, $(t/U)_c=0.059\,729\,1(8)$.}~\label{2dbh_jc}
\end{figure}

We extend the applicability of wrapping probability approach to the 2D unitary-filling BH model (\ref{HamBH}), aiming to precisely locate the U(1) QCP. We simulate the model in canonical ensemble with the worm QMC method within the imaginary-time path-integral representation. The simulations are performed on periodic $L\times L$ lattices with $L=8$, $16$, $32$, $48$, $64$, $80$, $96$ and $112$ for the temperature contour $\beta=2L$. Particle line wrapping probabilities $R_x$, $R_a$ and $R_2$ together with the SF stiffness $\rho_s$ are sampled. Figure~\ref{2dbh_jc} illustrates these quantities around the QCP. Least squares fits are performed for the finite-size scaling analyses described in Sec.~\ref{fss}. It is observed that the sampled wrapping probabilities $R_x$, $R_a$ and $R_2$ all exhibit slight finite-size corrections with the amplitudes of leading correction $|b_1| \lessapprox 0.04$. By contrast, the scaled SF stiffness $\rho_s L$ demonstrates more significant corrections with the amplitude $|b_1| \approx 0.2$. The observation of small corrections for the wrapping probabilities is reminiscent of that for the aforementioned 3D classical models. The fitting results are summarized in Appendix~\ref{appendix:2dbh} by which we estimate the critical hopping amplitude as $(t/U)_c=0.059\,729\,1(8)$.

\subsection{Determining Critical Exponents $\nu$ and $\eta$}\label{fatc}
Estimating critical exponents by performing fits at the precise critical temperature $T_c$ takes the advantage of reducing a fitting parameter. We perform extensive simulations with a number of lattice sizes ($L=12$, $16$, $20$, $24$, $32$, $40$, $48$, $64$, $80$, $96$, $128$, $160$, $192$, $256$, $384$ and $512$) right at the high-precision critical temperature $T_c=0.333\,067\,039$ of the 3D Villain model. For a quantity $\mathcal{Q}$ ($\mathcal{Q}=G_{R_xE}$, $G_{R_aE}$, $G_{R_2E}$, $G_{\rho_sE}$, $T_w$), the scaling ansatz reduces to
\begin{equation}\label{fit_at_Tc}
\mathcal{Q}=L^{y_\mathcal{Q}}(\mathcal{Q}_0+\sum_m b_m L^{-\omega_m}),
\end{equation}
with $y_\mathcal{Q}=1/\nu$ for $G_{R_xE}$, $G_{R_aE}$ and $G_{R_2E}$, $y_\mathcal{Q}=-1+1/\nu$ for $G_{\rho_sE}$, and $y_\mathcal{Q}=2-\eta$ for $T_w$.
\begin{table}
 \begin{center}
 \caption{Fits of $G_{R_xE}$, $G_{R_aE}$, $G_{R_2E}$ and $G_{\rho_sE}$ to $(\ref{fit_at_Tc})$ at the estimated critical temperature $T_c=0.333\,067\,039$ of the 3D Villain model. The correction exponent $\omega_1=0.789$ is adopted.}
 \label{Tab:estimate_nu}
 \begin{tabular}[t]{|l|l|l|l|l|l|}
 \hline
   Qua. & $L_{\rm min}$& $\chi^2/$DF  & $1/\nu$   & $\mathcal{Q}_0$  & $b_1$   \\
 \hline
  {\multirow{12}{*}{$G_{R_xE}$}}
                          &16& 17.4/13 &1.488\,47(14)&0.989\,0(5)&-  \\
                          &20& 16.8/12 &1.488\,53(17)&0.988\,8(7)&-  \\
                          &24& 14.5/11 &1.488\,38(20)&0.989\,4(8)&-   \\
                          & 32& 14.4/10 &1.488\,33(24)&0.990(1)&-  \\
                          & 40& 11.3/9 &1.488\,66(31)&0.988(1)&-   \\
                          & 48& 9.2/8 &1.488\,45(34)&0.989(2)&-   \\
                          & 16& 17.4/12 &1.488\,49(54)&0.989(3)&0.001(13) \\
                          & 20& 16.4/11 &1.488\,13(64)&0.991(3)&-0.01(2)   \\
                          & 24& 14.3/10 &1.488\,67(75)&0.988(4)&0.01(2)  \\
                          & 32& 13.5/9 &1.489\,2(10)&0.985(5)&0.03(3)   \\
                          & 40& 11.0/8 &1.488\,0(12)&0.992(7)&-0.03(5)   \\
                          & 48& 9.1/7 &1.488\,9(14)&0.987(8)&0.02(6) \\
 \hline
  {\multirow{6}{*}{$G_{R_aE}$}}
                           &16& 15.0/12 &1.487\,82(62)&1.344(4)&0.10(2) \\
                           &20& 13.5/11 &1.487\,32(74)&1.348(5) &0.08(3) \\
                           &24& 10.9/10 &1.488\,01(85)&1.343(6) &0.11(4) \\
                           &32& 10.3/9 &1.488\,5(11)&1.339(8)&0.14(5)  \\
                           &40& 8.7/8 &1.487\,5(14)&1.35(1)&0.07(8)  \\
                           &48& 8.4/7 &1.487\,9(15)&1.34(1) &0.1(1)  \\
 \hline
 {\multirow{6}{*}{$G_{R_2E}$}}
                         &  16& 16.7/12 &1.489\,18(72)&0.768(3)&0.10(1)   \\
                         &  20& 16.5/11 &1.488\,96(86)&0.769(4)&0.09(2)  \\
                         &  24& 14.9/10 &1.489\,6(10)&0.766(4)&0.11(2)  \\
                         &  32& 11.8/9 &1.491\,1(13)&0.760(6)&0.16(4)   \\
                         &  40& 10.0/8 &1.489\,6(17)&0.766(8)&0.10(6) \\
                         &  48& 9.9/7 &1.489\,9(19)&0.765(9)&0.11(7) \\

  \hline
  {\multirow{6}{*}{$G_{\rho_sE}$}}
                           &16& 16.7/12 &1.488\,42(54)&1.825(5)&-0.52(2) \\
                           &20& 16.7/11 &1.488\,33(65)&1.826(6) &-0.53(3)\\
                           &24& 15.2/10 &1.488\,83(77)&1.821(8) &-0.49(5) \\
                           &32& 14.9/9 &1.489\,2(10)&1.82(1)&-0.47(7) \\
                           &40& 13.0/8 &1.488\,1(12)&1.83(1)&-0.6(1)  \\
                           &48& 10.6/7 &1.489\,0(14)&1.82(1)&-0.5(1)  \\
 \hline
 \end{tabular}
 \end{center}
 \end{table}

\begin{figure}
\includegraphics[width=8cm,height=6cm]{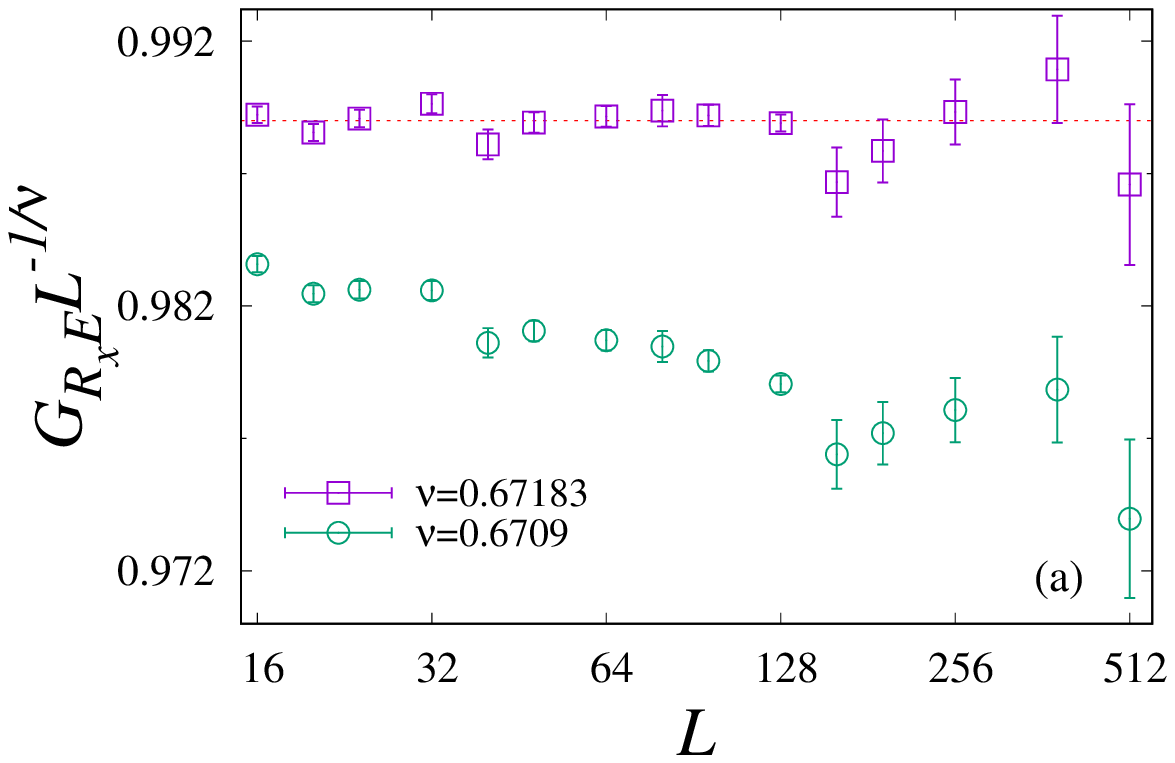}
\includegraphics[width=8cm,height=6cm]{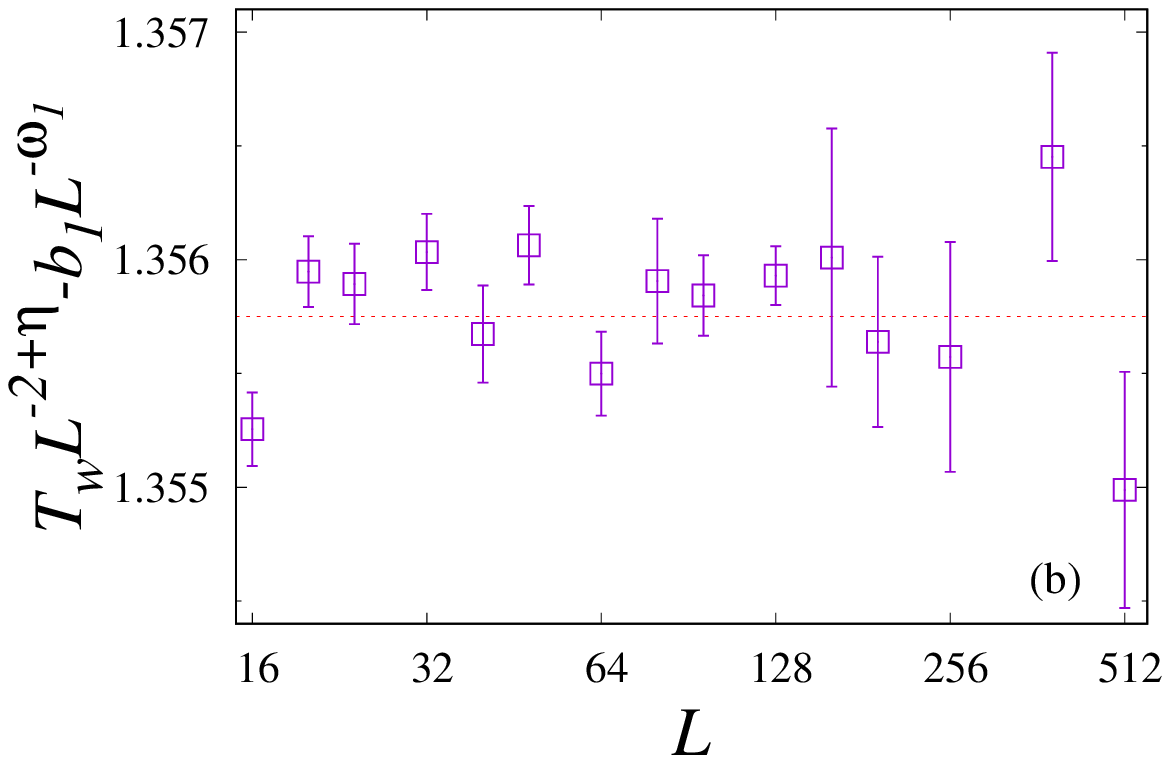}
\caption{(a) Scaled quantity $G_{R_xE} L^{-1/\nu}$ with $\nu=0.671\,83$ (estimate of this work) and $\nu=0.670\,9$ (experimental result~\cite{lipa2003specific}) at the estimated critical temperature $T_c=0.333\,067\,039$ of the 3D Villain model. The horizonal dashed line denotes $\mathcal{Q}_0=0.989$ which is determined from Table~\ref{Tab:estimate_nu}. (b) Scaled quantity $T_{w} L^{-2+\eta}$ with $\eta=0.038\,53$. A correction term quoted from Table~\ref{Tab:fit_Tw_Tc} is included. The horizonal dashed line denotes $\mathcal{Q}_0=1.355\,75$ which is determined from Table~\ref{Tab:fit-Villian2}.}~\label{Figs_Rxp}
\end{figure}

\begin{table}
 \begin{center}
 \caption{Fits of $T_w$ to $(\ref{fit_at_Tc})$ at the estimated critical temperature $T_c=0.333\,0670\,39$ of the 3D Villain model. The correction exponent $\omega_2=1.77$ is adopted. Some of the fits are performed with fixed $\mathcal{Q}_0=1.355\,75$ which is determined from Table~\ref{Tab:fit-Villian2}.}
  \label{Tab:fit_Tw_Tc}
 \begin{tabular}[t]{|l|l|l|l|l|l|l|}
 \hline
 $L_{\rm min}$& $\chi^2/$DF & $\eta$ & $\mathcal{Q}_0$  & $b_1$ &$\omega_1$ & $b_2$ \\
 \hline
                         20&13.3/11 &0.038\,59(16)&1.356(1)&-0.288(7)&0.789&- \\
                         24&13.3/10 &0.038\,59(19)&1.356(1)&-0.289(9)&0.789 &-\\
                         32&13.2/9 &0.038\,55(23)&1.356(2)&-0.29(1)&0.789&-\\
                         48&11.6/7 &0.038\,60(31)&1.356(3)&-0.29(2) &0.789&-\\
 \hline
                         12&17.7/12 &0.038\,45(23)&1.355(2)&-0.27(2)&0.789 &-0.23(9)\\
                         16&16.8/11 &0.038\,29(28)&1.354(2)&-0.26(2)&0.789&-0.4(2) \\
                         20&13.3/10 &0.038\,65(34)&1.357(3)&-0.29(3)&0.789&0.1(3) \\
                         24&13.2/9 &0.038\,68(40)&1.357(3)&-0.30(4) &0.789&0.1(4)\\
  \hline
                         20&14.5/11 &0.038\,53(4)&1.355\,75&-0.29(2)&0.791(20)&- \\
                         24&14.5/10 &0.038\,53(5)&1.355\,75&-0.29(2)&0.790(27)&- \\
                         32&14.3/9 &0.038\,51(7)&1.355\,75&-0.27(3) &0.776(39)&-\\
                         48&12.8/7 &0.038\,52(9)&1.355\,75&-0.27(7)&0.776(69)&- \\
  \hline
                         12&18.8/12 &0.038\,51(7)&1.355\,75&-0.25(4) &0.760(46)&-0.3(2)\\
                         16&17.0/11 &0.038\,45(9)&1.355\,75&-0.19(5) &0.694(68)&-0.7(3)\\
                         20&14.5/10 &0.038\,53(9)&1.355\,75&-0.3(1) &0.790(94)&0.0(6)\\
                         24&14.5/9 &0.038\,53(11)&1.355\,75&-0.3(1) &0.79(12)&0.0(9)\\
 \hline
 \end{tabular}
 \end{center}
 \end{table}

\begin{table}
 \begin{center}
 \caption{Fits of the wrapping probabilities $R_x$, $R_a$, $R_2$ and the scaled SF stiffness $\rho_s L$ to $(\ref{fit_at_Tc})$ with $y_\mathcal{Q}=0$ at the estimated critical temperature $T_c=0.333\,067\,039$ of the 3D Villain model. The $\mathcal{Q}_0$'s are fixed at their estimates determined from Table~\ref{Tab:fit-Villian1}. The correction exponent $\omega_2=1.77$ is adopted.}
 \label{Tab:fit-Villiannew-omega}
 \begin{tabular}[t]{|l|l|l|l|l|l|l|}
 \hline
  Qua.  &$L_{\rm min}$ & $\chi^2/$DF  & $\mathcal{Q}_0$ & $b_1$& $\omega_1$ & $b_2$  \\
 \hline
 {\multirow{4}{*}{$R_x$}}
                           &16& 4.6/12&0.378\,685&-0.034(6)&0.723(40)&-0.10(5) \\
                           &20& 2.9/11&0.378\,685&-0.04(1)&0.769(53)&-0.02(9) \\
                           &24& 2.9/10&0.378\,685&-0.04(2)&0.777(69)&0.0(1) \\
                           &32&  2.9/9&0.378\,685&-0.05(2)&0.78(10)&0.0(3) \\
\hline
 {\multirow{4}{*}{$R_a$}}
                           &16&  4.9/12&0.688\,920&-0.025(8)&0.702(71)&0.01(7) \\
                           &20&  2.2/11&0.688\,920&-0.04(2)&0.804(95)&0.2(1) \\
                           &24&  1.9/10&0.688\,920&-0.03(2)&0.76(12)&0.1(2) \\
                           &32&  1.8/9&0.688\,920&-0.04(4)&0.79(18)&0.1(4) \\
 \hline
 {\multirow{4}{*}{$R_2$}}
                           &16&  8.6/12&0.264\,021&-0.016(4)&0.647(59)&-0.16(4) \\
                           &20&  8.5/11&0.264\,021&-0.015(6)&0.635(77)&-0.18(7) \\
                           &24&  8.3/10&0.264\,021&-0.013(6)&0.603(96)&-0.21(9) \\
                           &32& 7.8/9&0.264\,021&-0.009(6)&0.52(14)&-0.3(2) \\
 \hline
 {\multirow{4}{*}{$\rho_s L$}}
                           &16&  8.5/12&0.515\,565&-0.13(1)&0.767(22)&-0.10(9) \\
                           &20&  5.3/11&0.515\,565&-0.16(2)&0.801(30)&0.1(2) \\
                           &24&  5.1/10&0.515\,565&-0.17(3)&0.812(38)&0.2(3) \\
                           &32&  4.4/9&0.515\,565&-0.21(6)&0.852(59)&0.6(5) \\
 \hline
 \end{tabular}
 \end{center}
 \end{table}

We estimate the critical exponent $\nu$ from the derivatives of wrapping probabilities $G_{R_\kappa E}$ ($\kappa=x, a, 2$) and the derivative of SF stiffness $G_{\rho_sE}$ by fitting them to (\ref{fit_at_Tc}). The details of the least squares fits are given in Tables~\ref{Tab:estimate_nu}. We find that the amplitudes of leading corrections for $G_{R_xE}$, $G_{R_aE}$ and $G_{R_2E}$ are typically smaller than that for $G_{\rho_sE}$, even though leading corrections are clearly there for $G_{R_aE}$ and $G_{R_2E}$. Moreover, it is hard to detect a finite amplitude of leading correction for $G_{R_xE}$, which should be very small if non-zero. This is a useful feature for locating $\nu$ as the leading correction term can be precluded out for the less uncertainties of fitting parameters. By fitting $G_{R_xE}$ to (\ref{fit_at_Tc}) without correction term, we determine $\nu=0.671\,83(6)$ for $L_{\rm min}=16$, $\nu=0.671\,80(8)$ for $L_{\rm min}=20$, $\nu=0.671\,87(9)$ for $L_{\rm min}=24$, $\nu=0.671\,89(11)$ for $L_{\rm min}=32$, $\nu=0.671\,75(14)$ for $L_{\rm min}=40$ and $\nu=0.671\,84(15)$ for $L_{\rm min}=48$, with $\chi^2/{\rm DF} \approx 1$ for all of these fits. Note that the fitting is already stable as $L_{\rm min} \gtrapprox 16$. On this basis, we estimate, more or less conservatively, that $\nu=0.671\,83(18)$. As an illustrative test of our estimate of $\nu$, we plot in Fig.~\ref{Figs_Rxp}(a) the scaled quantity $G_{R_xE}L^{-1/\nu}$ (without correction term) at $T_c$ with $\nu=0.671\,83$ and find that $G_{R_xE} L^{-1/\nu}$ converges fast. By contrast, the experimental estimate $\nu=0.670\,9$~\cite{lipa2003specific} is ruled out. Hence, our estimate of $\nu$ is further evidenced. For $G_{R_aE}$, $G_{R_2E}$ and $G_{\rho_sE}$, correction terms are needed to achieve a stable fitting, bringing about more uncertainties for fitting parameters.

In order to estimate $\eta$, we fit the finite-size $T_w$ data to (\ref{fit_at_Tc}) with $y_\mathcal{Q}=2-\eta$. Obvious corrections to scaling are present in the finite-size scaling. We explore the situations with different combinations of correction terms and the situations with $\mathcal{Q}_0$ being fixed or unfixed. By comparing all the fitting results listed in Table~\ref{Tab:fit_Tw_Tc}, our final estimate is $\eta=0.038\,53(48)$. The scaled quantity $T_{w} L^{-2+\eta}$ with a leading correction term is shown in Fig.~\ref{Figs_Rxp}(b) for $\eta=0.038\,53$.

On the basis of the fits for $T_w$ (Table~\ref{Tab:fit_Tw_Tc}), the leading correction exponent is estimated as $\omega_1=0.77(13)$, which is consistent with the literature results $\omega_1=0.789(11)$~\cite{guida1998critical} and $\omega_1=0.785(20)$~\cite{campostrini2006theoretical}.
Besides, we estimate the leading correction exponent $\omega_1$ from the finite-size scaling of quantities $R_x$, $R_a$, $R_2$ and $\rho_s L$, according to (\ref{fit_at_Tc}), with $y_\mathcal{Q}=0$. As shown by Table~\ref{Tab:fit-Villiannew-omega}, we obtain $\omega_1 \approx 0.7$ which agrees with the estimate $\omega_1=0.77(13)$ from analyzing $T_w$.

\section{Summary}~\label{sum}
In this work we utilize the geometric wrapping probability to exploit the quantitative aspects of the U(1) criticality in three dimensions in the contexts of the finite-temperature transitions in Villain and XY models and the quantum phase transition in the BH model. For both the classical and quantum models, we observe that certain wrapping probability-related quantities exhibit weak corrections in the finite-size scaling. The critical temperatures of the 3D XY and Villain models are estimated as $T_c=2.201\,844\,1(5)$ and $0.333\, 067\, 04(7)$, respectively. The QCP of the 2D unitary-filling BH model is estimated as $(t/U)_c=0.059\,729\,1(8)$. As demonstrated by Table~\ref{LitCP}, our locations of critical points for the 3D XY, 3D Villain and 2D unitary-filling BH models significantly improve over the best literature results. For the 3D classical models, the universal critical wrapping probabilities are determined as $R^c_x=0.378\,7(2)$, $R^c_a=0.688\,9(4)$ and $R^c_2=0.264\,0(3)$, which have not yet been reported. The critical winding number fluctuations is estimated as $\rho^c_sL=0.515\,6(3)$, which agrees well with $\rho^c_s L=0.516\,0(6)$ reported in~\cite{burovski2006high} and has a better precision. Moreover, we find that the derivative of a wrapping probability with respect to $T$, namely $G_{R_xE}$, suffers from negligible corrections. Making use of this feature, we determine the correlation length critical exponent as $\nu=0.671\,83(18)$, which is comparable with the most precise results available in literature (Table~\ref{Litnueta}). In addition, we estimate the critical exponent $\eta$ as $\eta=0.038\,53(48)$, which is close to the recent conformal bootstrap result $\eta=0.038 \, 52(64)$~\cite{kos2016precision}. To sum up, this work is a reference for applying wrapping probability-related quantities to determine the quantitative aspects of critical behaviors and provides several benchmarks for the 3D U(1) criticality.

\begin{acknowledgments} This work was supported by the National Natural Science Foundation of China under grant No. 11774002 (W.X., Y.S. and J.P.L.) and the National Science Fund for Distinguished Young Scholars under Grant No. 11625522 (Y.D.).
\end{acknowledgments}

\appendix
\section{Finite-size scaling for the thermodynamic phase transition of 3D XY model}~\label{appendix:3dxy}
The finite-size scaling of the dimensionless quantities $R_x$, $R_a$, $R_2$ and $\rho_sL$ for the 3D XY model are detailed in Table~\ref{Tab:fit-3dxy1}. For each quantity, the least squares fit with a leading correction term is performed. In some cases, a sub-leading correction term is further included. The leading correction amplitudes are found to be $|b_1|\approx 0.03$, $0.03$ and $0.02$ for wrapping probabilities $R_x$, $R_a$ and $R_2$, respectively. For $\rho_s L$, we observe more significant corrections with the amplitude $|b_1|\approx 0.1$. By comparing the fitting results of the dimensionless quantities, our final estimate of critical temperature is $T_c=2.201\,844\,1(5)$.

In Table~\ref{Tab:fit-3dxy1-fixnu}, we present the fitting results of the dimensionless quantities with $\nu$ being fixed at our final estimate $\nu=0.671\,83$. These fits produce an estimate of $T_c$ as $2.201\,844\,1(6)$, which is consistent with $T_c=2.201\,844\,1(5)$ obtained by the fits without a prior knowledge of $\nu$ (Table~\ref{Tab:fit-3dxy1}).

\begin{table*}
\begin{center}
\caption{Fits of the wrapping probabilities $R_x$, $R_a$, $R_2$ to $(\ref{fit_R})$ and the scaled SF stiffness $\rho_s L$ to $(\ref{fit_Rhos})$ for the 3D XY model. The correction exponents $\omega_1=0.789$ and $\omega_2=1.77$ are adopted.}
 \label{Tab:fit-3dxy1}
 \begin{tabular}[t]{|l|l|l|l|l|l|l|l|l|}
 \hline
  Qua. &$L_{\rm min}$ & $\chi^2/$DF  & $T_c$   & $1/\nu$   & $\mathcal{Q}_0$   & $a_1$   & $b_1$   & $b_2$   \\
 \hline
{\multirow{7}{*}{$R_x$}}
                             & 16&  28.4/31 & 2.201\,844\,12(13)&1.58(7)&0.378\,70(5) &0.06(2)& -0.030\,7(7)  & -   \\
                             & 32&  26.8/26 & 2.201\,844\,12(15)&1.59(8)&0.378\,69(8) &0.06(2)& -0.031(2)  & -   \\
                             & 64&  25.3/21 & 2.201\,844\,05(19)&1.57(8)&0.378\,8(1) &0.06(2)& -0.033(4)  & -   \\
                             &128&  16.9/15 & 2.201\,844\,00(33)&1.60(9)&0.378\,8(4) &0.05(2)& -0.04(2) & -   \\
                             & 8 &  30.1/35 & 2.201\,844\,01(14)&1.59(8)&0.378\,80(7) &0.05(2)& -0.034(1) & 0.041(9) \\
                             & 16&  28.4/30 & 2.201\,844\,11(17)&1.58(8)&0.378\,7(1) &0.06(2)& -0.031(3) & 0.00(4) \\
                             & 32&  26.3/25 & 2.201\,844\,01(23)&1.58(8)&0.378\,8(2) &0.06(2)& -0.037(9) & 0.1(2) \\
 \hline
 {\multirow{7}{*}{$R_a$}}
                             & 16&  31.8/31 & 2.201\,844\,47(13)&1.59(8)&0.688\,61(8) &0.07(3) & -0.015(1)  & -   \\
                             & 32&  26.8/26 & 2.201\,844\,32(15)&1.62(8)&0.688\,8(1) &0.06(3) & -0.019(2)  & -   \\
                             & 64&  24.4/21 & 2.201\,844\,16(20)&1.61(8)&0.689\,0(2) &0.07(3) & -0.027(6)  & -   \\
                             &128&  17.2/15 & 2.201\,844\,15(34)&1.63(9)&0.689\,0(5) &0.06(3) & -0.03(2)  & -   \\
                             & 8 &  31.8/35 & 2.201\,844\,07(14)&1.62(8)&0.689\,1(1) &0.06(3)& -0.033(2) & 0.21(1) \\
                             & 16&  27.1/30 & 2.201\,844\,23(17)&1.61(8)&0.688\,9(2) &0.06(3)& -0.026(5) & 0.12(6) \\
                             & 32&  25.1/25 & 2.201\,844\,09(24)&1.61(8)&0.689\,2(3) &0.07(3)& -0.04(1) & 0.4(3) \\
 \hline
 {\multirow{7}{*}{$R_2$}}
                             & 16&  38.6/31 & 2.201\,844\,05(17)&1.62(10)&0.264\,09(6) &0.04(2) & -0.023\,7(7)  & -   \\
                             & 32&  30.0/26 & 2.201\,844\,25(20)&1.60(10)&0.263\,97(8) &0.04(2) & -0.021(2)  & -   \\
                             & 64&  24.4/21 & 2.201\,843\,98(27)&1.57(10)&0.264\,2(1) &0.05(3) & -0.027(4)  & -   \\
                             &128&  16.0/15 & 2.201\,844\,27(45)&1.63(11)&0.263\,9(4) &0.04(2) & -0.01(2)  & -   \\
                             & 8 &  40.8/35 & 2.201\,844\,08(18)&1.61(10)&0.264\,07(7) &0.04(2)& -0.023(1) & -0.002(9) \\
                             & 16&  35.9/30 & 2.201\,844\,29(22)&1.60(10)&0.263\,9(1) &0.04(2)& -0.018(3) & -0.06(4) \\
                             & 32&  27.8/25 & 2.201\,843\,89(32)&1.59(10)&0.264\,3(2) &0.04(2)& -0.04(1) & $0.3(2)$ \\
 \hline
 {\multirow{7}{*}{$\rho_s L$}}
                             & 16&  32.9/31 & 2.201\,844\,19(14)&1.58(8)&0.515\,6(1) &0.11(5) & -0.098(1)  & -   \\
                             & 32&  26.7/26 & 2.201\,844\,13(16)&1.59(8)&0.515\,7(2) &0.10(4) & -0.100(3)  & -   \\
                             & 64&  23.9/21 & 2.201\,843\,92(22)&1.59(8)&0.516\,0(3) &0.10(5) & -0.112(8)  & -   \\
                             &128&  15.9/15 & 2.201\,844\,15(35)&1.68(10)&0.515\,5(7) &0.06(3) & -0.09(3)  & -   \\
                             & 8 &  33.4/35 & 2.201\,843\,98(15)&1.59(8)&0.515\,9(1) &0.10(5)& -0.109(3) & 0.14(2) \\
                             & 16&  31.8/30 & 2.201\,844\,07(18)&1.59(8)&0.515\,8(2) &0.10(5)& -0.104(7) & 0.08(8) \\
                             & 32&  24.7/25 & 2.201\,843\,84(26)&1.58(8)&0.516\,2(4) &0.10(5)& -0.13(2) & 0.5(4) \\
 \hline
 \end{tabular}
 \end{center}
 \end{table*}

\begin{table*}
\begin{center}
\caption{Fits of the wrapping probabilities $R_x$, $R_a$, $R_2$ to $(\ref{fit_R})$ and the scaled SF stiffness $\rho_s L$ to $(\ref{fit_Rhos})$ for the 3D XY model. The critical exponent $\nu$ is fixed at our final estimate $0.671\,83$ for a consistent check of the fits with unfixed $\nu$ (Table~\ref{Tab:fit-3dxy1}). The correction exponents $\omega_1=0.789$ and $\omega_2=1.77$ are adopted.}
 \label{Tab:fit-3dxy1-fixnu}
 \begin{tabular}[t]{|l|l|l|l|l|l|l|l|l|}
 \hline
  Qua. &$L_{\rm min}$ & $\chi^2/$DF  & $T_c$   & $1/\nu$   & $\mathcal{Q}_0$   & $a_1$   & $b_1$   & $b_2$   \\
 \hline
{\multirow{7}{*}{$R_x$}}
                             & 16&  30.0/32 & 2.201\,844\,13(13)&1/0.67183&0.378\,68(5) &0.095(4) & -0.0305(7)  &-   \\
                             & 32&  28.5/27 & 2.201\,844\,15(16)&1/0.67183&0.378\,66(8) &0.095(4) &-0.030(2) & -   \\
                            & 64&  26.6/22 & 2.201\,844\,05(21)&1/0.67183&0.378\,7(1) &0.095(4) &-0.033(4) &-   \\
                             &128&  18.6/16 & 2.201\,843\,92(38)&1/0.67183&0.378\,9(4) &0.094(4)& -0.04(2) & -   \\
                             & 8 &  31.8/36 & 2.201\,844\,02(15)&1/0.67183&0.378\,77(7) &0.094(4)& -0.034(1) & 0.040(9) \\
                             & 16&  30.0/31 & 2.201\,844\,14(18)&1/0.67183&0.378\,7(1) &0.095(4)& -0.030(3) & 0.00(4) \\
                             & 32&  27.9/26 & 2.201\,844\,00(25)&1/0.67183&0.378\,8(2) &0.094(4)& -0.04(1) & 0.1(2) \\
 \hline
 {\multirow{7}{*}{$R_a$}}
                             & 16&  33.6/32 & 2.201\,844\,52(14)&1/0.67183&0.688\,57(8) &0.132(5) & -0.015(1) & -   \\
                             & 32&  29.4/27 & 2.201\,844\,37(17)&1/0.67183&0.688\,7(1) &0.131(5) & -0.019(2)  & -   \\
                             & 64&  26.6/22 & 2.201\,844\,17(22)&1/0.67183&0.689\,0(2) &0.130(6) & -0.027(6)  & -   \\
                             &128&  19.5/16 & 2.201\,844\,07(40)&1/0.67183&0.689\,1(6) &0.130(6) & -0.03(2)  & -   \\
                             & 8 &  34.5/36 & 2.201\,844\,08(16)&1/0.67183&0.689\,1(1) &0.129(5)& -0.032(2) & 0.21(1) \\
                             & 16&  29.6/31 & 2.201\,844\,27(19)&1/0.67183&0.688\,9(2) &0.130(5)& -0.025(5) & 0.11(6) \\
                             & 32&  27.4/26 & 2.201\,844\,08(27)&1/0.67183&0.689\,2(3) &0.130(6)& -0.04(1) & 0.4(3) \\
 \hline
 {\multirow{7}{*}{$R_2$}}
                             & 16&  40.3/32 & 2.201\,844\,05(18)&1/0.67183&0.264\,07(6) &0.076(4) & -0.0235(7)  & -   \\
                             & 32&  31.3/27 & 2.201\,844\,29(21)&1/0.67183&0.263\,94(8) &0.077(4) & -0.020(2)  & -   \\
                             & 64&  25.2/22 & 2.201\,843\,99(29)&1/0.67183&0.264\,2(2) &0.076(4) & -0.027(5)  & -   \\
                             &128&  17.5/16 & 2.201\,844\,23(53)&1/0.67183&0.263\,9(4) &0.076(4) & -0.02(2)  & -   \\
                             & 8 &  42.5/36 & 2.201\,844\,10(20)&1/0.67183&0.264\,05(7) &0.076(4)& -0.023(1) &-0.003(9) \\
                             & 16&  37.2/31 & 2.201\,844\,35(24)&1/0.67183&0.263\,9(1) &0.077(4)& -0.017(3) & -0.07(4) \\
                             & 32&  28.9/26 & 2.201\,843\,87(35)&1/0.67183&0.264\,3(2) &0.076(4)& -0.04(1) & 0.3(2) \\
 \hline
 {\multirow{7}{*}{$\rho_s L$}}
                             & 16&  34.2/32 & 2.201\,844\,20(14)&1/0.67183&0.515\,5(1) &0.178(7) &-0.097(1)  & -   \\
                             & 32&  28.4/27 & 2.201\,844\,14(17)&1/0.67183&0.515\,6(2) &0.177(7) &-0.099(3)  & -   \\
                             & 64&  25.3/22 & 2.201\,843\,89(24)&1/0.67183&0.516\,0(3) &0.175(7) &-0.112(9)  & -   \\
                             &128&  19.9/16 & 2.201\,843\,96(43)&1/0.67183&0.515\,9(8) &0.174(8) &-0.10(3) & -   \\
                             & 8 &  34.9/36 & 2.201\,843\,97(16)&1/0.67183&0.515\,9(1) &0.176(7)&-0.109(3) & 0.13(2) \\
                             & 16&  33.3/31 & 2.201\,844\,07(20)&1/0.67183&0.515\,7(2) &0.177(7)&-0.103(7) & 0.07(8) \\
                             & 32&  26.0/26 & 2.201\,843\,79(29)&1/0.67183&0.516\,3(5) &0.175(7)&-0.13(2) & 0.6(4) \\
 \hline
 \end{tabular}
 \end{center}
 \end{table*}

\section{Finite-size scaling for the quantum phase transition of 2D unitary-filling BH model}~\label{appendix:2dbh}
The QCP of the 2D unitary-filling BH model is determined by finite-size scaling analyses of the dimensionless quantities $R_x$, $R_a$, $R_2$ and $\rho_sL$. The details of least squares fits are summarized in Table~\ref{Tab:fit-bh}. For each quantity, we perform fits without correction term or with different combinations of leading and sub-leading correction terms. It is observed that the leading correction amplitudes $|b_1|$ for the wrapping probabilities are $|b_1| \lessapprox 0.04$. For $\rho_s L$, it is found that $|b_1|\approx 0.2$. Note that the leading finite-size corrections of $R_x$, $R_a$ and $R_2$ are considerably smaller than that of $\rho_s L$. By comparing the fitting results in Table~\ref{Tab:fit-bh}, we provide the final estimate of QCP as $(t/U)_c=0.059\,729\,1(8)$.

We also perform fits with fixed critical exponent $\nu=0.671\,83$. The results are summarized in Table~\ref{Tab:fit-bh-fixnu}, which yields $(t/U)_c=0.059\,729\,1(7)$, agreeing with the analyses with unfixed $\nu$.

\begin{table*}
\begin{center}
\caption{Fits of the wrapping probabilities $R_x$, $R_a$, $R_2$ to $(\ref{fit_R})$ and the scaled SF stiffness $\rho_s L$ to $(\ref{fit_Rhos})$ for the 2D unitary-filling BH model. The correction exponents $\omega_1=0.789$ and $\omega_2=1.77$ are adopted.}
 \label{Tab:fit-bh}
 \begin{tabular}[t]{|l|l|l|l|l|l|l|l|l|}
 \hline
  Qua. &$L_{\rm min}$ & $\chi^2/$DF & $(t/U)_c$& $1/\nu$ & $\mathcal{Q}_0$  & $a_1$   & $b_1$& $b_2$ \\
  \hline
 {\multirow{6}{*}{$R_x$}}
               &48&31.0/28& 0.059\,728\,85(22)&1.39(13)& 0.205\,0(2)&-3(2)& -& -\\
               &64&25.9/20& 0.059\,729\,00(26)&1.39(16)& 0.205\,2(3)&-3(2)& -& -\\
               &16&34.2/43& 0.059\,729\,08(18)&1.464(44)& 0.206\,0(2)&-2.0(4)& -0.020(2)& -\\
               &32&32.1/35& 0.059\,729\,33(35)&1.452(87)& 0.206\,5(7)&-2.1(8)& -0.028(9)& -\\
                & 8& 37.4/50& 0.059\,729\,21(21)&1.467(38)&0.206\,3(3)&-2.0(3) &-0.027(4)&0.06(2)\\
               &16& 33.5/42& 0.059\,729\,40(43)&1.465(44)&0.207(1)&-2.0(4) &-0.04(2)&0.2(2)\\
 \hline
 {\multirow{6}{*}{$R_a$}}
                     &48&24.5/28& 0.059\,729\,01(23)&1.41(13)& 0.336\,1(3)&-4(2)&-& -\\
                     &64&20.3/20& 0.059\,729\,14(27)&1.40(16)& 0.336\,3(4)&-4(3)&-& -\\
                     &16&31.0/43&0.059\,728\,98(17)&1.467(41)&0.336\,6(3)&-3.0(5)&-0.016(3)&- \\
                     &32&27.1/35&0.059\,729\,40(34)&1.453(79)&0.338(1)&-3(1)&-0.03(1)&-\\
                     &8&32.4/50&0.059\,729\,23(20)&1.471(37)&0.337\,6(5)& -2.9(4)&-0.037(6)&0.18(3)\\
                     &16&29.0/42&0.059\,729\,50(41)&1.470(41)&0.339(2)& -2.9(5)&-0.06(3)&0.4(3)\\
 \hline
{\multirow{6}{*}{$R_2$}}
               &48&32.3/28& 0.059\,728\,25(35)&1.22(20)& 0.073\,9(2)&-3(2)& -& -\\
               &64&23.9/20& 0.059\,728\,64(41)&1.31(25)& 0.074\,1(2)&-2(2)& -& -\\
               &16&36.2/43& 0.059\,729\,28(27)&1.459(77)& 0.075\,3(2)&-1.0(3)&-0.023(2)& -\\
               &32&34.1/35& 0.059\,729\,03(54)&1.45(13)& 0.075\,0(5)&-1.1(6)&-0.020(7)& -\\
               & 8& 41.4/50& 0.059\,729\,04(32)&1.466(55)&0.074\,9(3)&-1.0(2) &-0.017(3)&-0.06(1)\\
               &16& 35.9/42& 0.059\,728\,99(66)&1.463(77)&0.074\,9(8)&-1.0(3) &-0.02(2)&-0.07(15)\\
 \hline
 {\multirow{6}{*}{$\rho_s L$}}
               &48&34.3/28& 0.059\,728\,88(28)&1.24(15)& 0.981(1)&-27(17)& -& -\\
               &64&28.4/20& 0.059\,729\,08(35)&1.19(18)& 0.982(2)&-34(27)& -& -\\
               &16&44.8/43& 0.059\,729\,12(19)&1.477(50)& 0.989(1)&-10(2)& -0.18(1)& -\\
               &32&41.1/35& 0.059\,729\,60(39)&1.429(86)& 0.994(4)&-12(4)& -0.25(5)& -\\
               & 8& 48.3/50& 0.059\,729\,30(23)&1.488(40)&0.992(2)&-9(2) &-0.24(2)&0.5(1)\\
               &16& 43.0/42& 0.059\,729\,65(45)&1.480(50)&0.996(6)&-10(2) &-0.3(1)&1(1)\\
 \hline
 \end{tabular}
 \end{center}
 \end{table*}

\begin{table*}
\begin{center}
\caption{Fits of the wrapping probabilities $R_x$, $R_a$, $R_2$ to $(\ref{fit_R})$ and the scaled SF stiffness $\rho_s L$ to $(\ref{fit_Rhos})$ for the 2D unitary-filling BH model. The critical exponent $\nu$ is fixed at our final estimate $0.671\,83$ for a consistent check of the fits with unfixed $\nu$ (Table~\ref{Tab:fit-bh}). The correction exponents $\omega_1=0.789$ and $\omega_2=1.77$ are adopted.}
 \label{Tab:fit-bh-fixnu}
 \begin{tabular}[t]{|l|l|l|l|l|l|l|l|l|}
 \hline
  Qua. &$L_{\rm min}$ & $\chi^2/$DF & $(t/U)_c$& $1/\nu$ & $\mathcal{Q}_0$  & $a_1$   & $b_1$& $b_2$ \\
  \hline
 {\multirow{6}{*}{$R_x$}}
               &48&31.6/29& 0.059\,728\,78(18)&1/0.67183& 0.205\,0(2)&-1.80(4)& -& -\\
               &64&26.2/21& 0.059\,728\,94(22)&1/0.67183& 0.205\,1(2)&-1.80(4)& -& -\\
               &16&34.5/44& 0.059\,729\,05(17)&1/0.67183& 0.205\,9(2)&-1.80(3)& -0.020(2)& -\\
               &32&32.2/36& 0.059\,729\,29(32)&1/0.67183& 0.206\,4(6)&-1.80(4)& -0.027(9)& -\\
                & 8& 37.7/51& 0.059\,729\,18(20)&1/0.67183&0.206\,3(3)&-1.81(3) &-0.027(4)&0.06(2)\\
               &16& 33.8/43& 0.059\,729\,37(41)&1/0.67183&0.207(1)&-1.80(3) &-0.04(2)&0.2(2)\\
 \hline
 {\multirow{6}{*}{$R_a$}}
                     &48&24.9/29& 0.059\,728\,94(18)&1/0.67183& 0.336\,0(3)&-2.71(6)& -& -\\
                     &64&20.6/21& 0.059\,729\,07(22)&1/0.67183& 0.336\,2(3)&-2.70(6)& -& -\\
                     &16&31.3/44&0.059\,728\,95(16)&1/0.67183&0.336\,5(3)&-2.72(5)&-0.016(3)&-\\
                     &32&27.3/36&0.059\,729\,35(32)&1/0.67183&0.337\,8(9)&-2.71(6)&-0.03(1)&-\\
                     &8&32.6/51&0.059\,729\,21(19)&1/0.67183&0.337\,6(5)& -2.72(5)&-0.036(6)&0.18(3)\\
                     &16&29.2/43&0.059\,729\,48(40)&1/0.67183&0.339(2)& -2.71(5)&-0.06(3)&0.4(3)\\
 \hline
{\multirow{6}{*}{$R_2$}}
               &48&34.1/29& 0.059\,728\,19(28)&1/0.67183& 0.073\,9(1)&-0.90(3)& -& -\\
               &64&24.4/21& 0.059\,728\,57(34)&1/0.67183& 0.074\,1(2)&-0.89(3)& -& -\\
               &16&36.3/44& 0.059\,729\,26(26)&1/0.67183& 0.075\,3(2)&-0.90(3)&-0.023(1)& -\\
               &32&34.2/36& 0.059\,728\,99(50)&1/0.67183& 0.075\,0(5)&-0.89(3)&-0.019(7)& -\\
               & 8& 41.6/51& 0.059\,729\,03(31)&1/0.67183&0.074\,9(2)&-0.90(3) &-0.017(3)&-0.06(1)\\
               &16& 36.1/43& 0.059\,728\,96(64)&1/0.67183&0.074\,8(8)&-0.90(3) &-0.02(2)&-0.1(1)\\
 \hline
 {\multirow{6}{*}{$\rho_s L$}}
               &48&37.1/29& 0.059\,728\,70(21)&1/0.67183& 0.980(1)&-9.4(2)& -& -\\
               &64&31.2/21& 0.059\,728\,86(24)&1/0.67183& 0.981(1)&-9.4(2)& -& -\\
               &16&44.8/44& 0.059\,729\,10(18)&1/0.67183& 0.989(1)&-9.4(2)& -0.18(1)& -\\
               &32&41.6/36& 0.059\,729\,52(35)&1/0.67183& 0.993(4)&-9.4(2)& -0.25(5)& -\\
               & 8& 48.3/51& 0.059\,729\,30(22)&1/0.67183&0.992(2)&-9.4(2) &-0.24(2)&0.5(1)\\
               &16& 43.0/43& 0.059\,729\,63(44)&1/0.67183&0.996(6)&-9.4(2) &-0.3(1)&1(1)\\
 \hline
 \end{tabular}
 \end{center}
 \end{table*}

\bibliography{papers}
\end{document}